\documentclass[aps,prd,reprint,superscriptaddress,amsmath,amssymb,longbibliography,nofootinbib]{revtex4-2}
\usepackage[T1]{fontenc}
\usepackage{lmodern,graphicx,bm,booktabs,array,xcolor,microtype}
\usepackage[colorlinks=true,linkcolor=blue!50!black,citecolor=blue!50!black,urlcolor=blue!50!black]{hyperref}
\newcommand{\MeV}{\,\mathrm{MeV}}
\newcommand{\GeV}{\,\mathrm{GeV}}
\newcommand{\keV}{\,\mathrm{keV}}
\newcommand{\half}{\tfrac12}
\newcommand{\threehalf}{\tfrac32}
\newcommand{\Om}{\Omega}
\newcommand{\dd}{\mathrm{d}}
\begin{document}
\raggedbottom
\title{Chiral partners of strange doubly heavy baryons:\\ Mass relations and decay predictions after the 2026 LHCb discoveries}
\author{Maciej A. Nowak}
\affiliation{Institute of Theoretical Physics and Mark Kac Center for Complex Systems Research,\\
Jagiellonian University, Kraków, Poland}
\author{Ismail Zahed}
\affiliation{Center for Nuclear Theory, Department of Physics and Astronomy,
Stony Brook University, Stony Brook, New York 11794, USA}
\date{9 October 2026}
\begin{abstract}
Three spectacular LHCb measurements in 2026---the observations of $\Omega_{cc}^{+}$, $\Xi_{cc}^{+}$, and the beauty-strange structure $B_{s0}^{*}(5700)^0$---make possible an experimentally constrained analysis of chiral partners of strange doubly heavy baryons. The doubly charmed masses test the earlier Ma--Harada prediction and support a compact, internally S-wave $cc$ color source. We combine this comparison with the heavy-quark--diquark hyperfine relation and with chiral gaps from the theory of chiral doublers, formulated in 1992 by Nowak, Rho and Zahed, to construct consistent $ccs$ and $bbs$ parity multiplets and a conditional extension to $bcs$. Under meson-to-baryon parity-gap transfer, the $349\MeV$ input from the $D_s$ sector gives negative-parity $ccs$ targets near $4.075$ and $4.183\GeV$. The separate $332\MeV$ input from the $B_s$ sector fixes the $bbs$ partner masses relative to their unknown ground-state mass, conditional on the new beauty-strange state's chiral-partner assignment. The primary mass predictions concern splittings; consistent quark-model and lattice estimates motivate an illustrative $\Omega_{bb}$ ground mass near $10.27\GeV$. We relate the differences of hyperfine splittings in the two parity sectors of mesons and baryons, and identify equal reduced single-pion widths, an E1 reduced-width ratio $1:8:4:5$, and a kaon selection rule whose leading predictions contain no new transition parameters. Absolute widths are estimated by matching the published Bardeen--Eichten--Hill and Harada--Rho--Sasaki meson calculations to a common convention. In the reference calculation, all members of the predicted $ccs$ and $bbs$ multiplets are narrow: the included decay channels give negative-parity widths of approximately $50$--$240\keV$, while the positive-parity spin partners are narrower still. In the proposed $\Omega_{QQ}\pi^0$ and $\Omega_{QQ}\gamma$ searches at LHCb, these states are expected to produce peaks dominated by detector resolution, allowing their masses and decay patterns to be investigated even when their intrinsic widths remain unresolved. The upper charm state is above kaon threshold but protected in the spectator limit. We provide a normalization audit, dipion phase-space integrals, mixed-flavor spin mixing, and reproducible calculations, while separating tests of heavy-source symmetry from evidence specific to chiral dynamics.
\end{abstract}
\maketitle
\section{Introduction}\label{sec:intro}
Heavy-light hadrons probe two complementary symmetry limits of QCD: heavy-quark spin symmetry for $m_Q\to\infty$~\cite{IsgurWise1989,IsgurWise1990,Georgi1990,ManoharWise} and light-quark chiral symmetry for $m_q\to0$. The proposal of Nowak, Rho and Zahed (NRZ)~\cite{NRZ}, first circulated in 1992, used these symmetries together to identify opposite-parity heavy-light multiplets as chiral partners. A similar realization of chiral doubling was subsequently developed independently by Bardeen and Hill~\cite{BH}. The central insight was that the parity splitting could encode spontaneous chiral symmetry breaking in the light cloud, while the heavy constituent supplies an approximately static color source.

Combining these limits is nontrivial. The heavy-quark spin decouples from the light dynamics and becomes a separately conserved quantum number. For massless light quarks, the chiral representations are organized by $q_{L,R}=(1\mp\gamma_5)q/2$, with helicity labeling massless one-particle states. A separately conserved constituent spin is not the appropriate counterpart of the static heavy spin. The NRZ effective action implements both symmetry structures, with a nonlinear realization of light chiral symmetry and a quark-loop derivation of the heavy-light interactions~\cite{NRZ}. Its $H=(0^-,1^-)$ and $G=(0^+,1^+)$ fields describe two heavy-spin doublets with the same light-cloud angular momentum $j_\ell=\half$ and opposite parity. Their mass difference is sensitive to the dynamically generated, chirality-breaking light-quark self-energy, whereas heavy-spin splittings within each doublet are suppressed by $1/m_Q$~\cite{NRZ2004}.

In the chiral-doubling construction, the opposite-parity multiplets arise from the left- and right-handed light-quark structure, with parity interchanging these components. Their separation reflects spontaneous chiral symmetry breaking and, in the NRZ construction, arises from the dynamically generated, chirality-breaking light-quark self-energy~\cite{NRZ,NRZ2004}. The partner is therefore introduced through chiral symmetry, rather than by adding a phenomenological orbital-excitation energy. The appropriate labels are the light-cloud angular momentum $j_\ell$ and parity: the two partners have the same $j_\ell$ and opposite parity.\footnote{In a nonrelativistic constituent-quark classification, $P_{q\bar q}=(-1)^{L+1}$ assigns the lowest $(0^+,1^+)$ channels to P waves. This correspondence does not explain their mass gap or require a relativistic light cloud to have a separately conserved $L$. We thank Marek Karliner for illuminating discussions of this distinction.} Their pion transition couplings are related to the parity splitting and the axial transition charge through the generalized Goldberger--Treiman relation implied by the axial Ward identity~\cite{NRZ2004}, with corrections away from the chiral and heavy-quark limits. Radiative transitions require additional electromagnetic matrix elements.

For doubly heavy baryons, replacing the heavy antiquark by a compact color-antitriplet diquark allows the same light-cloud picture to be tested. Heavy-quark--diquark symmetry (HDS) controls the leading spin relations~\cite{SavageWise,BVR,FM,HuMehen}; the additional chiral-partner hypothesis concerns the opposite-parity light clouds. This separation is essential: a hyperfine relation alone does not establish the chiral origin of a parity gap.

Three remarkable LHCb observations in 2026 supply the experimental motivation: $\Omega_{cc}^{+}$~\cite{LHCb}, $\Xi_{cc}^{+}$~\cite{Xi2026}, and a narrow beauty-strange structure near $5700\MeV$~\cite{Bs2026}. Together with $\Xi_{cc}^{++}$, discovered in 2017~\cite{Xi2017}, the baryon measurements test an earlier prediction for the strange light cloud. The beauty structure offers an additional parity-gap input, conditional on its spin-parity assignment. Its chiral interpretation was recently examined in Ref.~\cite{NZ2026}.

The low scalar masses illustrate why the chiral perspective matters. A representative relativized potential calculation gives $M(D_{s0}^*)=2484\MeV$~\cite{GodfreyMoats2016}, about $166\MeV$ above the observed $2317.7\MeV$~\cite{PDG}. Two potential-model implementations give $M(B_{s0}^*)=5831$ and $5805\MeV$~\cite{GodfreyMoatsSwanson2016}, respectively $132$ and $106\MeV$ above the new $5698.9\MeV$ structure if it is the scalar partner~\cite{Bs2026}. These discrepancies expose limitations of the specified valence potential spectra and motivate a treatment of the light cloud that incorporates chiral dynamics. They are not a universal statement about all quark models or a unique demonstration of chiral doubling.

Quantitative studies of heavy-light meson decays were subsequently developed by Bardeen, Eichten and Hill (BEH)~\cite{BEH} and Harada, Rho and Sasaki (HRS)~\cite{HRS}, using formalisms based on the chiral-doubling scenario of NRZ and Bardeen--Hill~\cite{NRZ,BH}. Both works calculate hadronic transitions, while BEH also estimates electromagnetic widths. Their hadronic decay calculations provide the two matching prescriptions compared below, which we combine with a common, separately specified electromagnetic model. BEH also discussed the extension to doubly heavy baryons, while Ma and Harada explicitly developed their chiral-partner spectrum and transitions~\cite{MaHarada,MaHarada2017}. The theoretical construction therefore predates the new observations; these measurements now allow its light-cloud relations to be confronted with data and its baryon targets to be specified more directly.

The new measurements supply previously unavailable experimental inputs for a joint analysis of the charm and bottom baryon spectra. The specific contributions of this work are:
\begin{enumerate}
\item \textbf{An experimental test of the common light cloud.} We confront the earlier Ma--Harada strange--nonstrange mass prediction with the new doubly charmed baryon masses, keeping the two $\Xi_{cc}$ charge states distinct and specifying what the agreement establishes about the compact-source picture (Sec.~\ref{sec:experiment}).
\item \textbf{Linked charm and bottom search targets.} We combine the measured charm ground-state mass, the separate $D_s$ and $B_s$ parity-gap inputs, and the $3/4$ hyperfine relation in both parity sectors. Spin centroids give consistent $ccs$ and $bbs$ chiral rectangles, with bottom predictions expressed primarily as splittings relative to the unknown ground state. The beauty input remains conditional on the new signal's partner assignment (Sec.~\ref{sec:spectrum}).
\item \textbf{A set of tests independent of transition normalization.} We relate the mesonic and baryonic differences of hyperfine splittings between parity sectors and identify equal reduced single-pion widths and the reduced E1 ratio $1:8:4:5$. We state which unknown quantities cancel and which symmetry assumptions remain, so that these tests can be used without fitting absolute decay matrix elements (Sec.~\ref{sec:robust}).
\item \textbf{A quantitative account of narrowness.} We resolve the charged and neutral kaon thresholds and show why the upper charm partner can remain narrow even above the lowest threshold: its S-wave channel is closed, while the open D-wave channel vanishes in the spectator limit. We quantify how light-cloud mixing can lift this suppression and give the condition for complete kaon closure in bottom (Sec.~\ref{sec:kaon}).
\item \textbf{A common comparison of meson-matched baryon widths.} We adapt the BEH and HRS inputs with consistent normalization, baryon kinematics and heavy-spin factors, and combine them with explicit electromagnetic and dipion benchmarks. The channel-resolved calculation and its sensitivity analysis separate robust ratios from model-dependent absolute widths (Secs.~\ref{sec:widths} and \ref{sec:results}, and the appendices).
\item \textbf{A conditional mixed-flavor extension.} We exhibit the $bcs$ spin-mixing problem and its consequences for parity gaps and coherent transition amplitudes, identifying the additional inputs needed to turn unmixed reference values into physical predictions (Sec.~\ref{sec:bc}).
\end{enumerate}
The novelty lies in this application to the new measurements and in the resulting quantitative spectrum and decay tests. The chiral-doubling construction, the $3/4$ hyperfine relation and the angular-momentum methods are established ingredients. Section~\ref{sec:tests} sets out experimental and lattice measurements that can test the resulting predictions and probe the chiral origin of the parity gaps.

The distinction between a symmetry relation and a numerical width is central. Heavy-spin ratios can test a light-cloud assignment without determining its chiral origin. Conversely, transferring a chiral gap does not fix an E1 radial matrix element. The analysis therefore states the assumptions of each result and retains the possible axial feed-down interpretation of the beauty signal.

\section{Experimental inputs and the compact-source test}\label{sec:experiment}
The reported central masses are $3725.9$, $3619.97$, and $5698.9\MeV$, for $\Omega_{cc}^{+}$, $\Xi_{cc}^{+}$ and the beauty-strange structure, respectively. Ma and Harada~\cite{MaHarada2017} predicted in 2017
\begin{equation}\label{eq:historical}
 M_{\Omega_{cc}}-M_{\Xi_{cc}}\simeq104.9\MeV,
 \qquad M_{\Omega_{cc}}\simeq3726.3\MeV.
\end{equation}
Their input was the $\Xi_{cc}^{++}$ mass measured at the time. With the precision reference $M_{\Xi_{cc}^{++}}=3621.55\MeV$~\cite{PDG}, the new strange--nonstrange difference is $104.35\MeV$; using the new $\Xi_{cc}^{+}$ gives $105.93\MeV$. The agreement is striking. It supports the picture in which soft light degrees of freedom see an approximately pointlike color-antitriplet $cc$ source, with no resolved internal orbital excitation. The assumed source has $L_{cc}=0$ and $S_{cc}=1$, not zero angular momentum altogether. Mass agreement alone neither measures its radius nor uniquely determines its orbital wavefunction. The precision and isospin conventions are detailed in Appendix~\ref{app:experiment}.

The beauty structure provides a complementary mass input. Under direct $0^+\to B_s^0\pi^0$ decay, its mass implies a $332.0\MeV$ parity gap. LHCb also allows an axial state near $5748.3\MeV$ decaying through $B_s^{*0}\pi^0$ with an unobserved photon, or overlapping scalar and axial contributions~\cite{Bs2026}. Its width limit is $9.8\MeV$ at 90\% confidence, not a measured keV partial width. We retain the assignment ambiguity. The scalar interpretation and its predicted axial partner were discussed in Ref.~\cite{NZ2026}; the present objective is the linked baryon spectrum and its testable transition pattern.

An explicit numerical prediction for the beauty-strange parity splitting was given by Nowak, Rho and Zahed in their 2004 paper~\cite{NRZ2004}, first circulated in 2003. With the strange-quark mass input $m_s=150\MeV$ used in that calculation, they predicted the chiral partners of $B_s$ and $B_s^*$ to lie $323\MeV$ above the corresponding ground states. This is only about $9\MeV$ below the $332\MeV$ splitting inferred, under the direct scalar interpretation, from the beauty-strange signal announced by LHCb on 1 August 2026~\cite{Bs2026,BsAnnouncement2026}. The close numerical agreement provides a test of that earlier mesonic prediction; the present analysis uses the updated $332\MeV$ input for the $bbs$ sector.

The experimental ground-state assignments are used here as the lowest $J^P=\half^+$ baryons. The mass agreement does not itself supply a direct spin-parity determination. The charm flavor test and the beauty candidate therefore play different roles in the argument: the former constrains the ground-state light cloud; the latter motivates the bottom parity-gap benchmark.

\section{Heavy-source symmetry and the reference spectrum}\label{sec:spectrum}
\subsection{State assignments and assumptions}
The states considered here contain a ground-state heavy diquark and a strange light cloud. For identical heavy quarks, the Pauli principle requires $S_{QQ}=1$. Coupling this spin to $j_\ell=\half$ gives
\begin{equation}
 B_J=\Om_{QQ}(J^+),\qquad X_J=\Om_{QQ}(J^-),\qquad J=\half,\threehalf .
\end{equation}
The negative-parity pair is assigned to the opposite-parity strange light cloud around the same internally unexcited heavy diquark. States involving an internal diquark excitation need not belong to this pair. A $j_\ell=\threehalf$ multiplet has different light-cloud angular momentum and can mix with $X_{3/2}$ at finite heavy mass.

Our working assumptions are: a compact spectator diquark; leading heavy-spin symmetry for transition amplitudes; the specified meson-like chiral gaps; and a common constituent-model radial matrix element for E1 transitions. These assumptions do not determine all low-energy constants from the gaps alone. In particular, a mass gap does not by itself fix an electromagnetic width. The numbers below are a reproducible realization of these assumptions, not parameter-free consequences of chiral symmetry.

\subsection{Hyperfine matching and parity centroids}

Let $\delta_\pm=M_{3/2}^{\pm}-M_{1/2}^{\pm}$. The Brambilla--R\"osch--Vairo and Fleming--Mehen analyses~\cite{BVR,FM} imply, for the same heavy flavor and a common light cloud,
\begin{equation}\label{eq:hf}
\begin{aligned}
 \delta_+&=\frac34\delta_H(1+\epsilon_Q),\\
 \delta_H&=M_{Q\bar s}^{1^-}-M_{Q\bar s}^{0^-}.
\end{aligned}
\end{equation}
Here $\epsilon_Q\equiv4\delta_+/(3\delta_H)-1$ is the dimensionless relative correction to the leading HDS relation for the positive-parity baryon pair; the reference calculation sets $\epsilon_Q=0$. The ground-state ratio receives a relative perturbative correction of order $\alpha_s^2$. Mehen and Mohapatra~\cite{MehenMohapatra} argue that its nonperturbative corrections scale as $\Lambda_{\rm QCD}^2/m_Q^2$, rather than $\Lambda_{\rm QCD}/m_Q$, within their compact-diquark expansion. Finite-size effects and the extension to an excited light cloud require separate control; we do not assign their ground-state uncertainty estimate to both parity sectors. Allowing a separate relative correction for the opposite-parity light cloud, we write
\begin{equation}\label{eq:hfnegative}
\begin{aligned}
 \delta_-&=\frac34\delta_G(1+\epsilon_{Q,-}),\\
 \delta_G&=M_{Q\bar s}(1^+)-M_{Q\bar s}(0^+).
\end{aligned}
\end{equation}
Here $\epsilon_{Q,-}$ need not equal $\epsilon_Q$. Both are set to zero when quoting leading HDS predictions. Appendix~\ref{app:masses} derives the $3/4$ factor from chromomagnetic matching and spin algebra. The short-distance chromomagnetic coefficient cancels in the leading same-flavor ratio. The relation constrains a \emph{hyperfine} splitting; it supplies no factor $3/4$ for the chiral gap. Using the rounded meson splittings $143.8$ and $48.5\MeV$~\cite{PDG} gives
\begin{equation}
 \delta_{cc}=107.85\MeV,\qquad \delta_{bb}=36.375\MeV.
\end{equation}
The experimental input errors alone would give approximately $0.30$ and $1.05\MeV$, respectively, but Eq.~\eqref{eq:hf} has larger, unquantified symmetry corrections, particularly for charm.

We use the spin centroid
\begin{equation}\label{eq:centroid}
 \overline M_\pm=\frac{M_{1/2}^{\pm}+2M_{3/2}^{\pm}}{3},\qquad
 \Delta_\chi=\overline M_- -\overline M_+,
\end{equation}
and transfer the spin-independent excitation under the common-light-cloud partner hypothesis,
\begin{equation}\label{eq:gaptransfer}
 \Delta_\chi\simeq\bar M_G-\bar M_H,\qquad
 \bar M_{H,G}=\frac{M_0+3M_1}{4}.
\end{equation}
Kinetic, finite-size and coupled-channel effects can modify this transfer; chiral symmetry alone does not enforce its exact numerical value. For the reference calculation we adopt the equal-hyperfine approximation $\delta_- =\delta_+$. The specified gaps are then the same in both spin channels:
\begin{equation}\label{eq:gap}
\begin{gathered}
 \Delta_\chi^{cc}=349\MeV,\qquad\Delta_\chi^{bb}=332\MeV,\\
 M(X_J)=M(B_J)+\Delta_\chi.
\end{gathered}
\end{equation}
The $332\MeV$ bottom gap is now supported by the $5698.9\MeV$ beauty-strange structure, conditional on a direct scalar decay to $B_s^0\pi^0$~\cite{Bs2026,NZ2026}. LHCb also permits axial feed-down from a state near $5748.3\MeV$, or overlapping scalar and axial signals. Thus the numerical gap remains conditional on the partner assignment; it is not taken from the known $j_\ell=\threehalf$ bottom-strange excitations. Equal hyperfine splittings in the two parity sectors are a further approximation; Appendix~\ref{app:masses} gives the general formula and a charm refinement.

\subsection{Illustrative absolute bottom mass}\label{sec:bottommass}
Our principal spectroscopic predictions concern hyperfine and parity splittings. An absolute $\Omega_{bb}$ mass additionally requires the change in heavy-diquark binding between charm and bottom, which is not fixed by the chiral gap or the $3/4$ relation. To give an illustrative mass scale, we use two distinct theoretical approaches: the Karliner--Rosner quark-model calculation~\cite{KarlinerRosner2018} and lattice QCD~\cite{Brown2014}.

The quark-model extrapolation must be performed between spin centroids. With $\Delta_{\rm cent}=\overline M_{\Omega_{bb}}-\overline M_{\Omega_{cc}}$, Eq.~\eqref{eq:centroid} gives
\begin{equation}\label{eq:absolute-bottom}
 M_{\Omega_{bb}}=M_{\Omega_{cc}}+\Delta_{\rm cent}
                 +\frac23(\delta_{cc}-\delta_{bb}).
\end{equation}
Here $M_{\Omega_{cc}}=3725.90\MeV$ and the last term is $47.65\MeV$. Retaining the spin-independent binding increments of the two Karliner--Rosner constituent-mass prescriptions, while using our measured charm normalization and hyperfine matching, gives $M_{\Omega_{bb}}=10260.8$ and $10271.6\MeV$. These are our charm-calibrated extrapolations, not the original absolute predictions of Ref.~\cite{KarlinerRosner2018}; their bookkeeping is given in Appendix~\ref{app:masses}. Independently, Brown, Detmold, Meinel and Orginos~\cite{Brown2014} obtain
\begin{equation}
 M_{\Omega_{bb}}^{\rm lattice}=10273(27)(20)\MeV,
\end{equation}
with statistical and systematic errors, respectively.

The numerical consistency of these approaches motivates the illustrative choice $M_b\equiv M_{\Omega_{bb}}=10270\MeV$ in Table~\ref{tab:masses} and the recoil factors. This is neither a measurement nor an absolute-mass prediction of chiral doubling. A provisional search interval $10.20$--$10.35\GeV$ expresses the possible scale of additional theory uncertainty; it is not a confidence interval or an average of the quoted calculations. The two quark-model prescriptions share assumptions, and agreement with one lattice calculation does not establish a few-MeV accuracy. The relative bottom spectrum remains applicable when a different ground-state mass is adopted.

\begin{table*}[t]
\centering
\caption{Mass inputs and resulting benchmark spectrum, in MeV. The $ccs$ ground-state mass is the reported LHCb central value~\cite{LHCb}; the spin assignment is assumed here. The illustrative $bbs$ ground mass is motivated by the consistent quark-model and lattice estimates discussed in Sec.~\ref{sec:bottommass}. The primary predictions concern splittings; excited masses are conditional on Eq.~\eqref{eq:gap}. Digits specify the calculation, not absolute-mass accuracy.}
\begin{tabular}{lrrrrr}
\toprule
System & $\Delta_\chi$ & $B_{1/2}$ & $B_{3/2}$ & $X_{1/2}$ & $X_{3/2}$\\
\midrule
$ccs$ &349&3725.90&3833.75&4074.90&4182.75\\
$bbs$ &332&10270.00&10306.38&10602.00&10638.38\\
\bottomrule
\end{tabular}
\label{tab:masses}
\end{table*}

Absolute masses enter recoil factors weakly, whereas proximity to a kaon threshold can be decisive. We therefore specify the ground-state mass inputs explicitly rather than attach an artificial precision to unmeasured levels.

The charm masses are useful search targets. For bottom, the symmetry statements are more cleanly expressed as offsets $(0,36.375,332,368.375)\MeV$ from $M(B_{1/2})$, ordered as $(B_{1/2},B_{3/2},X_{1/2},X_{3/2})$. The illustrative absolute bottom mass in Table~\ref{tab:masses} locates this relative pattern and sets the recoil factors. Its uncertainty is largely common to all four masses; separate corrections affect the predicted splittings. The gap construction does not determine the overall mass uncertainty.

\begin{figure*}[t]
\centering
\includegraphics[width=\textwidth]{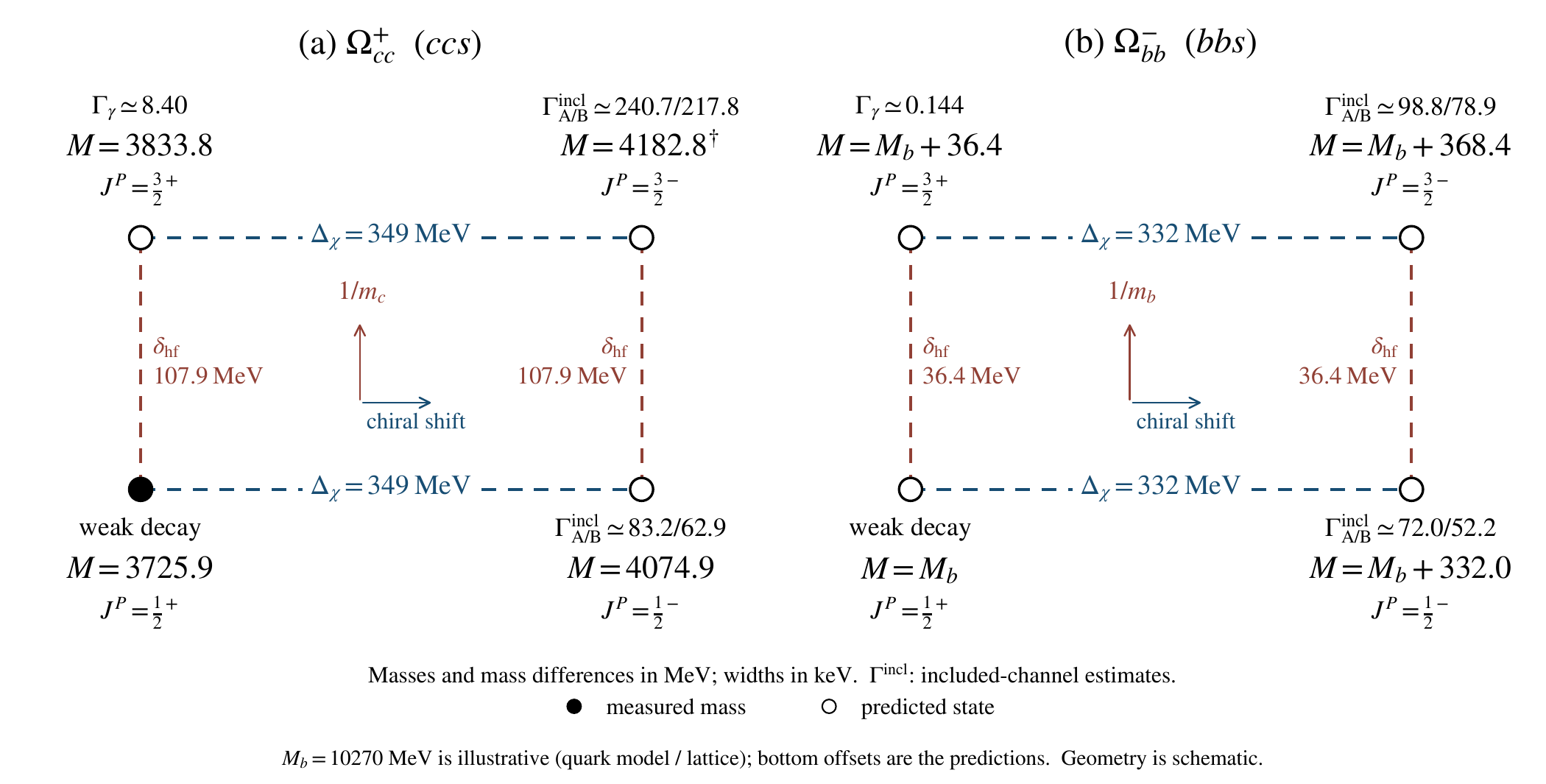}
\caption{Strange doubly heavy baryon chiral rectangles, following the meson
organization of Ref.~\cite{NZ2026}. The filled circle denotes the measured
charm ground-state mass, with an assumed $J^P=1/2^+$ assignment;
open circles denote predicted states.
The compact spin-one heavy diquark and $j_\ell=1/2$ light-cloud construction
uses equal parity-sector hyperfine splittings and the reference chiral
gaps shown; the bottom gap is conditional on the beauty-strange partner
interpretation. Widths above the masses are in keV. Negative-parity
$\Gamma_{\rm A/B}^{\rm incl}$ are the included-channel BEH-/HRS-matched
estimates of Eq.~\eqref{eq:scenarios}, with a common electromagnetic
overlap $R=2.5\,\mathrm{GeV}^{-1}$ and strange constituent mass
$m_s=0.323\GeV$, not complete totals or an uncertainty band. Positive-parity excited widths are M1 estimates; weak widths are
not calculated. The bottom ground mass $M_b$ is unmeasured; the widths use
$M_b=10270\MeV$ as an illustration motivated by consistent quark-model and
lattice estimates (Sec.~\ref{sec:bottommass}). The principal mass predictions
are the displayed offsets, independent of this illustrative absolute scale.
The dagger marks
the upper charm state: $\Xi_{cc}\bar K$ is kinematically open but forbidden in
the leading spectator limit; a mixing-induced contribution is omitted.
Bottom kaon closure is conditional on the unknown strange--nonstrange
splitting. Digits specify benchmarks, not theory precision;
parity-dependent hyperfine corrections deform the rectangles.}
\label{fig:baryonic-rectangles}
\end{figure*}

Figure~\ref{fig:baryonic-rectangles} summarizes the charm and bottom reference spectra as chiral rectangles, following the mesonic organization of Ref.~\cite{NZ2026}. The widths shown above the masses use the prescriptions developed in Sec.~\ref{sec:widths}; they are included-channel estimates rather than total widths.

\section{Predictions with the least model dependence}\label{sec:robust}
The following hierarchy separates symmetry tests from absolute-width models. ``Parameter free'' means no additional transition parameter at the stated leading order, after the state assignments and required mass inputs are specified.

\subsection{Differences of hyperfine splittings between parity sectors}
Define the baryon parity gaps by $\Delta_J=M(X_J)-M(B_J)$. To distinguish the mesonic and baryonic quantities, write
\begin{equation}\label{eq:defect}
 \begin{aligned}
 \mathcal D_Q^{\rm mes}&\equiv\delta_G-\delta_H,\\
 \mathcal D_Q^{\rm bar}&\equiv\Delta_{3/2}-\Delta_{1/2}=\delta_--\delta_+,\\
 \mathcal D_Q^{\rm bar}&\simeq\frac34\mathcal D_Q^{\rm mes}.
 \end{aligned}
\end{equation}
The second line is an exact mass identity. Indeed,
\[
\begin{aligned}
 \Delta_{3/2}-\Delta_{1/2}
 &=[M(X_{3/2})-M(B_{3/2})]\\
 &\quad-[M(X_{1/2})-M(B_{1/2})]\\
 &=[M(X_{3/2})-M(X_{1/2})]\\
 &\quad-[M(B_{3/2})-M(B_{1/2})].
\end{aligned}
\]
Thus the difference of the two baryon parity gaps is also the difference of the hyperfine splittings in the negative- and positive-parity sectors. The third line of Eq.~\eqref{eq:defect}, which relates this baryonic quantity to its mesonic counterpart, is the leading HDS prediction, not an exact identity. No subtraction of an absolute meson mass from an absolute baryon mass is involved. An exactly rectangular spectrum requires $\mathcal D_Q^{\rm bar}=0$.

The mesonic input quoted in the 2026 LHCb analysis~\cite{Bs2026} is $\mathcal D_c^{\rm mes}=-2.7\pm0.7\MeV$. Multiplication of both its central value and input uncertainty by $3/4$ gives
\[
 \mathcal D_c^{\rm bar}\simeq(-2.025\pm0.525)\MeV
 \simeq(-2.03\pm0.53)\MeV.
\]
The latter number is therefore the predicted \emph{baryonic} difference of hyperfine splittings. Its negative sign means that the negative-parity hyperfine interval, and equivalently the spin-$3/2$ parity gap, is smaller than its positive-parity or spin-$1/2$ counterpart by about $2\MeV$. The quoted uncertainty is propagated from the mesonic input only. With the corrections defined in Eqs.~\eqref{eq:hf} and \eqref{eq:hfnegative}, the additional contribution is
\[
 \mathcal D_Q^{\rm bar}-\frac34\mathcal D_Q^{\rm mes}
 =\frac34(\delta_G\epsilon_{Q,-}-\delta_H\epsilon_Q),
\]
which need not cancel or be small compared with this few-MeV difference. The spin-averaged parity gap cancels from the test: this relation probes HDS but does not alone establish the chiral origin of that gap.

\subsection{Equal reduced S-wave pion widths} In the isospin limit, $I=0\to0+1$ forbids $X_J\to B_J\pi^0$. Isospin breaking induces a leading spin-diagonal amplitude. For
\begin{equation}\label{eq:phi}
 \Phi_J=\frac{M(B_J)}{M(X_J)}E_{\pi,J}^{2}p_{\pi,J},
\end{equation}
heavy-spin symmetry gives
\begin{equation}\label{eq:pionratio}
 \frac{\Gamma(X_{1/2}\to B_{1/2}\pi^0)}{\Phi_{1/2}}
 =\frac{\Gamma(X_{3/2}\to B_{3/2}\pi^0)}{\Phi_{3/2}}.
\end{equation}
The unknown transition coupling and $\eta$--$\pi^0$ mixing cancel. This equality is more robust than either absolute pion width. Cross-spin pion transitions vanish at this leading order.

\subsection{An E1 ratio without a radial overlap} A rank-one light-cloud operator coupled to a spectator spin one has fractions
\begin{equation}\label{eq:robustweights}
 w_{J_iJ_f}=\begin{pmatrix}1/9&8/9\\4/9&5/9\end{pmatrix}.
\end{equation}
Rows and columns run over $\half,\threehalf$. Appendix~\ref{app:angular} derives each entry explicitly from the Wigner--Eckart theorem. Therefore, defining $\widehat\Gamma_{if}=\Gamma_{E1}(X_i\to B_f\gamma)/k_{if}^{3}$,
\begin{equation}\label{eq:e1ratio}
 \boxed{\widehat\Gamma_{11}:\widehat\Gamma_{13}:
 \widehat\Gamma_{31}:\widehat\Gamma_{33}=1:8:4:5.}
\end{equation}
Here subscripts 1 and 3 denote twice the spin. The unknown reduced E1 matrix element cancels, as do the common charge and wavefunction factors. The unpolarized widths include their initial-spin average. The relation is for E1 partial widths, not total radiative widths with arbitrary higher multipoles. Heavy-spin-breaking currents and admixtures produce corrections. Together with Eq.~\eqref{eq:pionratio}, it tests the assumed light-cloud assignment independently of the absolute-width normalization; neither relation alone uniquely diagnoses chiral doubling.

The scalar pion operator is diagonal with unit weight; the vector operator yields Eq.~\eqref{eq:robustweights}. The same rank argument excludes a leading D-wave kaon transition between pure $j_\ell=\half$ clouds, as developed in Sec.~\ref{sec:kaon}.

\subsection{Conditional parity-gap transfer}
Let $d_0=M(0^+)-M(0^-)$ and $d_1=M(1^+)-M(1^-)$ denote the two meson parity gaps, so that $d_1-d_0=\mathcal D_Q^{\rm mes}$. The meson centroid weights $1:3$ give
\[
 \bar M_G-\bar M_H=\frac{d_0+3d_1}{4}
 =d_0+\frac34\mathcal D_Q^{\rm mes}.
\]
The baryon centroid weights $1:2$ instead imply the exact identities
\[
\begin{aligned}
 \Delta_{1/2}&=\Delta_\chi-\frac23\mathcal D_Q^{\rm bar},\\
 \Delta_{3/2}&=\Delta_\chi+\frac13\mathcal D_Q^{\rm bar},
\end{aligned}
\]
as derived in Appendix~\ref{app:masses}. Substituting the centroid-transfer assumption, Eq.~\eqref{eq:gaptransfer}, and the leading HDS relation, $\mathcal D_Q^{\rm bar}\simeq(3/4)\mathcal D_Q^{\rm mes}$, gives
\begin{align}\label{eq:resolved}
 \Delta_{1/2}&\simeq M(0^+)-M(0^-)+\frac14(\delta_G-\delta_H),\nonumber\\
 \Delta_{3/2}&\simeq M(1^+)-M(1^-).
\end{align}
The coefficient $1/4$ in the first line is $3/4-(2/3)(3/4)$; the second line uses $3/4+(1/3)(3/4)=1$. These coefficients follow from the different meson and baryon spin weights, not from a universal rescaling of the chiral gap by $3/4$. There is no new fitted baryon gap, but the common-cloud and partner hypotheses are additional physical assumptions. The distinction between the rounded $349\MeV$ reference gap and an exact meson-centroid input is displayed numerically in Appendix~\ref{app:masses}.

Under the direct scalar interpretation of the beauty structure and equal meson hyperfine intervals, the axial target is
\begin{equation}
 M(B_{s1}')\simeq5698.9+48.5=5747.4\MeV.
\end{equation}
The input-only uncertainty is approximately $2.1\MeV$, with additional theory and assignment uncertainties. The observed peak could already contain this axial member through feed-down~\cite{Bs2026,NZ2026}.

\subsection{Hierarchy of assumptions}
The exact statements are isospin conservation in its symmetry limit and kinematic closure below a physical threshold. The $3/4$ relations, reduced transition ratios and D-wave null test are leading heavy-source or heavy-spin statements. The baryon parity centroids additionally require the partner assignment and the meson-to-baryon transfer. Absolute E1, M1, one-pion and dipion widths require matrix elements or model matching. This hierarchy is essential when interpreting the numerical results below.

\section{Absolute decay-width benchmarks}\label{sec:widths}
\subsection{Kinematics and single-pion matching}
For a decay into a baryon of mass $M_f$ and a particle of mass $m$, define
\begin{equation}\label{eq:kin}
\begin{aligned}
 p(M_i,M_f,m)&=\frac{\sqrt{\lambda(M_i^2,M_f^2,m^2)}}{2M_i},\\
 E_\pi&=\frac{M_i^2-M_f^2+m_{\pi^0}^2}{2M_i},
\end{aligned}
\end{equation}
where $\lambda(x,y,z)=(x-y-z)^2-4yz$ and $p=0$ below threshold. The isospin-violating, leading S-wave pion decay has the form
\begin{equation}\label{eq:pi}
\begin{aligned}
 \Gamma(X_J\to B_J\pi^0)&=C_\pi\Phi(M_i,M_f),\\
 \Phi(M_i,M_f)&=\frac{M_f}{M_i}E_\pi^2p(M_i,M_f,m_{\pi^0}).
\end{aligned}
\end{equation}
The spin sum is identical for $J=\half$ and $\threehalf$. Cross-spin pion decays vanish at this order. Isospin forbids $I=0\to0+1$ in the symmetry limit; neutral-pion emission is induced by isospin breaking, usually represented by $\eta$--$\pi^0$ mixing.

Our comparison transfers reduced light-cloud amplitudes under heavy-quark--diquark symmetry. Appendix~\ref{app:transfer} explains baryon-number conservation, the heavier source, and the limitations of this transfer. BEH~\cite{BEH} quotes $21.5\keV$ for each strange charm single-pion channel at $G_A=1$. HRS~\cite{HRS} quotes $4.17\keV$ for the scalar channel and $3.75\keV$ for the axial channel. We match the scalar channel of each paper to Eq.~\eqref{eq:pi}:
\begin{align}\label{eq:match}
 C_\pi^{\rm A}&=\frac{21.5\times10^{-6}\GeV}{\Phi(2.3177,1.96835)}
 =7.9528\times10^{-4}\GeV^{-2},\\
 C_\pi^{\rm B}&=\frac{4.17\times10^{-6}\GeV}{\Phi(2.317,1.969)}
 =1.5591\times10^{-4}\GeV^{-2}.
\end{align}
Mass arguments in these formulas are in GeV. The HRS reference masses are those of its table. In this convention the pion-width ratio is $C_\pi^{\rm B}/C_\pi^{\rm A}=0.1960$. The matching is to \emph{published model estimates}, not measured meson partial widths. Applying the same reduced coefficient to a baryon invokes heavy-quark--diquark symmetry, with finite-mass and compositeness corrections left unresolved. Appendix~\ref{app:normalization} explains why direct substitution of baryon masses into the old meson formulas would not constitute a consistent normalization prescription.

\subsection{Electric and magnetic dipole transitions}
Treat the diquark as a point constituent with mass $M_d=m_Q+m_{Q'}$ and charge $e_d=e_Q+e_{Q'}$, in units of the positron charge. For the relative coordinate $\bm r=\bm r_s-\bm r_d$, the internal electric dipole is $e\,e_{\rm eff}\bm r$, where
\begin{equation}\label{eq:charge}
e_{\rm eff}=\frac{e_sM_d-e_dm_s}{M_d+m_s}.
\end{equation}
We use the effective constituent masses $m_s=0.323\GeV$, $m_c=1.530\GeV$ and $m_b=4.730\GeV$.\footnote{The strange constituent mass is chosen to be consistent with the $323\MeV$ bottom-strange chiral gap predicted by NRZ~\cite{NRZ2004}. Equating these numerical scales is a phenomenological matching assumption for the electromagnetic operators. The spectrum retains the updated $332\MeV$ meson-gap input. (Here, a short-distance strange-quark mass means a renormalized current mass.)} We retain the BEH~\cite{BEH} electromagnetic operator prescription and heavy constituent masses, while replacing its $m_s=0.480\GeV$ input [Eq.~(59) of that reference]. The resulting charges are $-0.42881$ for $ccs$, $-0.30032$ for $bbs$ and $-1/3$ for $bcs$. In particular, the charge cancellation that suppresses some charm-strange \emph{meson} E1 amplitudes does not transfer to $ccs$.

For a $1P,j_\ell=\half\to1S,j_\ell=\half$ light-cloud transition,
\begin{equation}\label{eq:E1}
\Gamma_{E1}(X_{J_i}\to B_{J_f}\gamma)=\frac{4\alpha}{9}e_{\rm eff}^2R^2k^3w_{J_iJ_f}.
\end{equation}
Here $k=(M_i^2-M_f^2)/(2M_i)$ and the reference overlap is $R=2.5\GeV^{-1}$, with $R=\int\dd r\,r^3u_{1S}(r)u_{1P}(r)$ for radial wavefunctions normalized with $r^2\dd r$. The angular weights are those of Eq.~\eqref{eq:robustweights}, with rows and columns ordered by spins $\half,\threehalf$.
Each row sums to unity. Different photon energies must be used for every channel: the energy releases are $\Delta_\chi$, $\Delta_\chi-\delta$, and $\Delta_\chi+\delta$, before recoil. Replacing all of them by the chiral gap would miss the principal enhancement of the upper charm state's radiative width.

The M1 widths within each parity doublet are evaluated with the same constituent model:
\begin{equation}\label{eq:M1}
\begin{aligned}
 \Gamma_{M1}(B_{3/2}\to B_{1/2}\gamma)&=\frac{4\alpha}{9}(a_S-a_d)^2k^3,\\
 \Gamma_{M1}(X_{3/2}\to X_{1/2}\gamma)&=\frac{4\alpha}{9}(a_P-a_d)^2k^3.
\end{aligned}
\end{equation}
The same $m_s=0.323\GeV$ is used in the strange magnetic moment and the orbital current. The magnetic coefficients, including the light P-wave orbital contribution, are given in Appendix~\ref{app:angular}. Equations~\eqref{eq:E1}--\eqref{eq:M1} use nonrelativistic constituent-state normalization with exact photon momentum and unit long-wavelength overlap for M1. A heavy-hadron EFT convention may introduce an additional $M_f/M_i$ factor~\cite{HuMehen}; this is a finite-recoil model choice, not an extra factor to mix into one channel selectively.

\subsection{Two-pion contributions and the meaning of the comparison}
Isospin conservation does \emph{not} forbid an isoscalar two-pion final state. The scalar-mediated amplitude is proportional to the dipion recoil momentum, so the decay is strongly suppressed close to threshold. We use the scalar propagator shape of BEH~\cite{BEH}, with masses $1.0$ and $1.5\GeV$, and normalize the summed meson dipion width to $4.2\keV$. We impose the isoscalar charged-to-neutral multiplicities $2:1$, including their different thresholds. This is a specified adaptation of the scalar model; Appendix~\ref{app:dipion} documents the normalization.

HRS~\cite{HRS} computes an isospin-violating $\phi$--$\rho$ vector-mixing contribution, quoting $2.13\times10^{-4}\keV$ for $D_{s1}\to D_s\pi^+\pi^-$. It explicitly leaves scalar exchange out of that result. A small vector contribution is therefore not evidence for a small \emph{complete} dipion width. Our two comparisons are
\begin{align}\label{eq:scenarios}
 \Gamma_{\rm A}^{\rm incl}&=\Gamma_{\rm EM}+\Gamma_{\pi}^{\rm A}+\Gamma_{\pi\pi}^{S},\\
 \Gamma_{\rm B}^{\rm incl}&=\Gamma_{\rm EM}+\Gamma_{\pi}^{\rm B}+\Gamma_{\pi\pi}^{V}.
\end{align}
Here $\Gamma_{\rm EM}$ includes E1 and, for $X_{3/2}$, the intradoublet M1 cascade. The superscript ``incl'' means \emph{included channels}. Scenario B is an HRS-matched hadronic subtotal combined with our common electromagnetic model, not a published HRS baryon total. Adding the same scalar benchmark to B provides the more direct comparison of single-pion normalizations. Scalar and vector dipion amplitudes have different exchange/isospin symmetry; their interference integrates to zero over the full symmetric charged-pion phase space at this order.

\section{Numerical results and sensitivity}\label{sec:results}
\begin{table*}[t]
\centering
\caption{Opposite-parity radiative partial widths in keV at $m_s=0.323\GeV$ and $R=2.5\GeV^{-1}$. All final baryons in the four E1 columns have positive parity. The last column is the negative-parity M1 cascade.}
\begin{tabular}{lrrrrr}
\toprule
 & $\half\to\half$ & $\half\to\threehalf$ & $\threehalf\to\half$ & $\threehalf\to\threehalf$ & $X_{3/2}\to X_{1/2}\gamma$\\
\midrule
$ccs$&15.44&42.46&133.46&77.46&1.843\\
$bbs$&7.09&40.25&38.54&35.45&0.00989\\
\bottomrule
\end{tabular}
\label{tab:em}
\end{table*}

\begin{table*}[t]
\centering
\caption{Calculated widths in keV. The scalar and vector entries sum over accessible positive-parity final spins and pion charges. A and B denote Eq.~\eqref{eq:scenarios}; kaon widths, additional levels and uncalculated operators are excluded. Digits beyond roughly two significant figures identify the benchmark, not its accuracy.}
\begin{tabular}{llrrrrrr}
\toprule
System &$J^P$&$\pi^0$ A&$\pi^0$ B&$\pi\pi$ scalar&$\pi\pi$ vector&A included&B included\\
\midrule
$ccs$&$\half^-$&25.33&4.97&0.00360&$1.25\,10^{-6}$&83.23&62.86\\
     &$\threehalf^-$&25.47&4.99&2.500&$1.63\,10^{-4}$&240.74&217.76\\
$bbs$&$\half^-$&24.69&4.84&0.00105&$6.37\,10^{-7}$&72.04&52.18\\
     &$\threehalf^-$&24.70&4.84&0.0654&$1.54\,10^{-5}$&98.77&78.85\\
\bottomrule
\end{tabular}
\label{tab:widths}
\end{table*}

For the upper $ccs$ state, adding the scalar benchmark to B changes $217.76$ to $220.26\keV$; it scarcely changes the other B entries. Thus the difference of the included widths is driven mainly by the single-pion normalization. The large ratio between the upper and lower charm widths arises from the $k^3$ enhancement of $X_{3/2}\to B_{1/2}\gamma$, not from assigning a larger chiral gap to the excited spin state.

The positive-parity $B_{3/2}$ states have no open pion channel at these hyperfine splittings. Their M1 widths are
\begin{equation}\label{eq:positive}
\begin{aligned}
 \Gamma[\Om_{cc}(\threehalf^+)\to\Om_{cc}(\half^+)\gamma]&=8.40\keV,\\
 \Gamma[\Om_{bb}(\threehalf^+)\to\Om_{bb}(\half^+)\gamma]&=0.1436\keV.
\end{aligned}
\end{equation}
The lowest $B_{1/2}$ states decay weakly. Their weak widths are not predicted by the present chiral calculation. Narrow electromagnetic or isospin-violating decays are nevertheless prompt on detector scales; ``narrow'' here refers to the natural resonance width.

\begin{figure*}[t]
\centering
\includegraphics[width=0.85\textwidth]{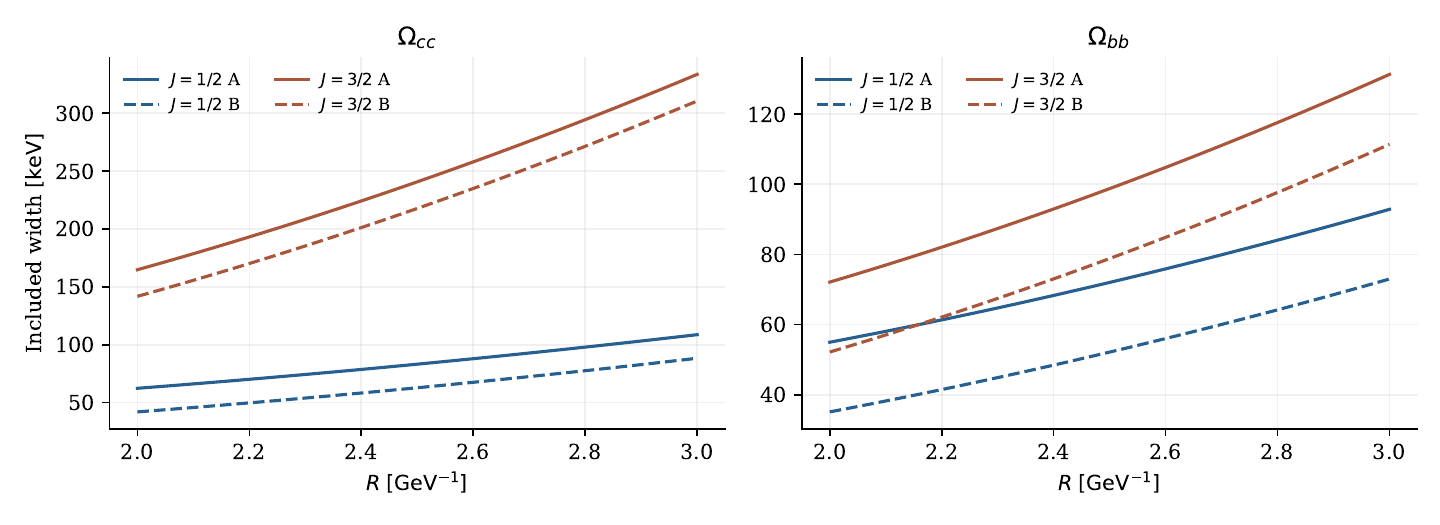}
\caption{Dependence of the included widths on the assumed E1 radial overlap. Solid curves: A; dashed curves: B. All other inputs are fixed. These curves are model sensitivities, not confidence intervals, and exclude the possible kaon contribution.}
\label{fig:sensitivity}
\end{figure*}

Varying $R$ from $2$ to $3\GeV^{-1}$ gives A ranges $(62\text{--}109,165\text{--}334)\keV$ for charm and $(55\text{--}93,72\text{--}131)\keV$ for bottom. The corresponding B ranges are $(42\text{--}88,142\text{--}311)$ and $(35\text{--}73,52\text{--}111)\keV$. These variations dominate the A--B distinction for the upper charm state. The quoted uncertainty $k_{\rm HRS}=0.59\pm0.07$ in the HRS meson calibration alone would imply about $24\%$ in its pion widths; it does not represent the full uncertainty of the baryon extrapolation. Scalar masses, diquark size, recoil, the light-cloud wavefunction and symmetry-breaking transition operators supply further uncertainties. We do not combine these into a fictitious statistical error bar.

The constituent-mass choice supplies a separate electromagnetic sensitivity. At fixed hadron masses and $R$, using $m_s=0.323\GeV$ instead of the original BEH value $0.480\GeV$ decreases every $ccs$ E1 width by $16.4\%$ and increases every $bbs$ E1 width by $11.0\%$. For $bcs$, $e_d=+1/3$ gives $e_{\rm eff}=-1/3$ independently of $m_s$, so its E1 widths are unchanged at fixed overlap. The strange magnetic moment grows in magnitude, increasing the M1 widths. With both electric and magnetic terms included, the negative-parity widths decrease by about $12$--$16\%$ for charm and increase by about $7$--$10\%$ for bottom across the A/B prescriptions. The meson-matched masses, thresholds and hadronic coefficients do not depend explicitly on this constituent input. This comparison holds the radial overlap fixed; its possible dynamical response requires a wavefunction calculation.

\section{Kaon thresholds and heavy-spin breaking}\label{sec:kaon}
The relevant strangeness-conserving channel is $\Om_{QQ}\to\Xi_{QQ}\bar K$. Define
\begin{equation}
 d_Q=M[\Om_{QQ}(\half^+)]-M[\Xi_{QQ}(\half^+)].
\end{equation}
If strange and nonstrange hyperfine splittings are approximately equal, the lowest diagonal S-wave kaon channels close for $\Delta_\chi+d_Q<m_K$. The stronger condition that \emph{every} ground-doublet kaon channel be kinematically closed is
\begin{equation}\label{eq:threshold}
 \Delta_\chi+\delta+d_Q<m_K.
\end{equation}
These conditions are different.

For charm, $M(\Xi_{cc}^{++})=3621.55\MeV$ and $m_{K^-}=493.677\MeV$~\cite{PDG} give a lowest threshold of $4115.227\MeV$. The lower state of Table~\ref{tab:masses} lies $40.33\MeV$ below it; the upper state lies $67.52\MeV$ above it. Using a nonstrange hyperfine splitting near $106.6\MeV$, the $\Xi_{cc}^*\bar K$ threshold remains about $39\MeV$ above $X_{3/2}$. Its principal S-wave kaon decay is therefore closed. The newly measured $\Xi_{cc}^{+}$ also fixes the neutral-kaon threshold at $4117.581\MeV$; both charge channels are retained below.

The residual $X_{3/2}\to\Xi_{cc}(\half^+)\bar K$ channel requires a D wave by total angular momentum and parity. With a spectator heavy spin it is forbidden for a pure $j_\ell=\half$ initial state: a rank-two operator cannot connect $j_\ell=\half$ to $\half$. This is a dynamical heavy-spin selection rule, not phase-space closure. A $j_\ell=\threehalf$ admixture, or an independent heavy-spin-breaking operator, can open the decay.

For illustration, define the admixture by $|X_{3/2}\rangle=\cos\theta\,|j_\ell=\half\rangle+\sin\theta\,|j_\ell=\threehalf\rangle$ and a D-wave coupling $g_T$ through
\begin{equation}\label{eq:Dwave}
 \Gamma_K=\sin^2\theta\sum_{a=K^-,\bar K^0}
 \frac{2g_T^2}{3\pi f_{132}^2\Lambda_\chi^2}\frac{M_{\Xi_a}}{M_X}
 p_a^5\,w_D,\qquad w_D=\frac15,
\end{equation}
where $f_{132}=\sqrt2(93.3\MeV)$ and $\Lambda_\chi=1\GeV$. This convention reproduces the usual meson coefficient $4g_T^2/(15\pi f_{132}^2\Lambda_\chi^2)$ for $D_2\to D\pi$~\cite{Batra}. Setting $g_T=0.4$ and using $M(\Xi_{cc}^{++})=3621.55\MeV$, $M(\Xi_{cc}^{+})=3619.97\MeV$~\cite{PDG,Xi2026}, and physical charged/neutral kaon masses yields
\begin{equation}
 \Gamma_K=0.616\MeV\,\sin^2\theta\left(\frac{g_T}{0.4}\right)^2.
\end{equation}
The additions are $4.68$, $18.58$ and $72.06\keV$ at $\theta=5^\circ,10^\circ,20^\circ$, respectively. Neither $g_T$ nor $\theta$ has been determined here; this is a sensitivity estimate for one mechanism and does not bound independent $1/m_Q$ operators.

For bottom, Eq.~\eqref{eq:threshold} requires $d_b<125.30\MeV$. An illustrative $d_b=90\MeV$ closes even the lowest kaon channel of the upper state by $35.30\MeV$. The conclusion is conditional on this flavor splitting and on corrections to the assumed gaps. The $\eta$ channels are closed for the entire benchmark because the largest intrafamily release is below $m_\eta$.

\section{Mixed-flavor extension: \texorpdfstring{$bcs$}{bcs}}\label{sec:bc}
There are two heavy-diquark spins, $S_{bc}=0,1$. Consequently, each parity sector contains \emph{two} $J=\half$ basis states and one $J=\threehalf$ state. The same-flavor $3/4$ rule cannot assign their physical spectrum without a mixing calculation. A useful spin Hamiltonian is
\begin{equation}\label{eq:bcham}
H_{\rm spin}=a\,\bm s_c\cdot\bm j_\ell+b\,\bm s_b\cdot\bm j_\ell+c\,\bm s_c\cdot\bm s_b.
\end{equation}
For the benchmark, $a=71.9\MeV$ and $b=24.25\MeV$. These values follow from the same point-diquark matching as Eq.~\eqref{eq:hf}; $c$ is an independent heavy-heavy spin coupling. In the $(S_{bc}=0,S_{bc}=1)$ basis,
\begin{equation}\label{eq:bcmat}
\begin{aligned}
 H_{1/2}&=\begin{pmatrix}-3c/4&\sqrt3(b-a)/4\\\sqrt3(b-a)/4&c/4-(a+b)/2\end{pmatrix},\\
 H_{3/2}&=c/4+(a+b)/4.
\end{aligned}
\end{equation}
The off-diagonal element is $-20.63\MeV$. Neglecting it gives a triplet-doublet hyperfine difference
\begin{equation}
 \delta_{bc}^{\rm unmixed}=\frac34(a+b)=\frac{\delta_{cc}+\delta_{bb}}2=72.1125\MeV.
\end{equation}
This is a basis-state benchmark, not a measured or symmetry-fixed physical separation.

For a transparent extension we take an unmixed triplet reference mass $M(B_{1/2})=7.000\GeV$ and $\Delta_{bc}=340.5\MeV$, the midpoint of the two assumed gaps. The interpolation is a modeling choice. With $R=2.5\GeV^{-1}$, the resulting quantities are given in Table~\ref{tab:bcwidths}.
\begin{table*}[t]
\centering
\caption{Conditional $S_{bc}=1$ basis-state results, in keV. They omit physical $S_{bc}=0$ mixing and transitions involving the additional state.}
\label{tab:bcwidths}
\begin{tabular}{lrrrrr}
\toprule
$J^P$&$\pi^0$ A&$\pi^0$ B&EM&A included&B included\\
\midrule
$\half^-$&25.81&5.06&45.83&71.64&50.89\\
$\threehalf^-$&25.84&5.07&110.97&137.35&116.03\\
\bottomrule
\end{tabular}
\end{table*}
The corresponding unmixed $B_{3/2}\to B_{1/2}\gamma$ width is $1.767\keV$. A separate unmixed $S_{bc}=0$ parity pair with the same $7.000\GeV$ ground-state mass has $\Gamma_{E1}=82.87\keV$ and pion widths $25.81$ or $5.06\keV$. Its included scalar-model A width is $108.70\keV$, and its vector-only B width is $87.93\keV$. The common scalar/triplet reference mass in this last illustration is not a solution of Eq.~\eqref{eq:bcmat}; it only compares reduced light-cloud transitions at fixed kinematics.

The mixing is potentially substantial. For example, $c=25\MeV$ gives $J=\half$ spin-energy eigenvalues $-53.93$ and $-6.65\MeV$, a $J=\threehalf$ energy $30.29\MeV$, and a triplet probability $0.744$ in the lower eigenstate. There is then no unique $72\MeV$ physical hyperfine interval. If the spin Hamiltonians and mixing angles are identical in the two parity sectors, a common additive chiral gap survives in the eigenbasis. Otherwise both the parity gaps and pion amplitudes change. Electromagnetic amplitudes must always be mixed coherently, as described in Appendix~\ref{app:bc}; the table cannot be converted into physical widths by averaging its entries. Even the kaon threshold is marginal for an illustrative $d_{bc}=90\MeV$: the unmixed upper state is about $8.94\MeV$ above $\Xi_{bc}\bar K$. We therefore make no unconditional total-width prediction for $bcs$.

\section{Experimental and lattice tests}\label{sec:tests}
The strongest result is a pattern: the lowest positive-parity states are weakly decaying, their spin partners are radiatively narrow, and the proposed light-cloud negative-parity states have small isospin-violating pion widths when the diagonal kaon channels close. With ordinary E1 overlaps, the latter widths are naturally of order $10^{-1}\MeV$, rather than universally a few keV. The bottom upper state has a more favorable complete kaon closure condition than the charm upper state.

The numbers do not establish the chiral origin of a level splitting. A lattice test should determine both parity sectors with operators for light-cloud and diquark excitations, extract their mixing, and include $\Xi_{QQ}\bar K$ operators wherever thresholds are relevant. The spin-averaged gaps and the two hyperfine splittings should be studied separately as light-quark masses and the heavy mass vary. Matrix elements of the light-quark axial current provide an additional test of the proposed partner assignment, as explained below. Near chiral restoration, parity-projected correlators and their symmetry transformations are the useful observables; a vacuum mass fit alone is insufficient.

For phenomenology, radiative branching ratios and photon energies can test the recoupling pattern after phase space is removed. Meson single-pion partial-width measurements would replace the model calibrations in Eq.~\eqref{eq:match}. A determination of the electromagnetic light-cloud matrix element and the $j_\ell=\threehalf$ admixture would reduce the two largest uncertainties. Distinguishing complete versions of the BEH and HRS dynamics would additionally require a common treatment of scalar exchange and higher-order operators; the vector-only dipion subtotal is not such a discriminator.

An experimentally useful sequence is to establish the positive-parity hyperfine partner, identify both negative-parity spins, and compare the extracted centroid gap and the difference of hyperfine splittings between parity sectors with their mesonic counterparts. The reduced E1 ratios require partial widths rather than uncorrected event yields; photon phase space, reconstruction efficiencies and possible cascade feed-down must be accounted for. A nonzero $X_{3/2}\to\Xi_{cc}\bar K$ width probes the size of mixing or independent heavy-spin-breaking operators. Failure of a leading relation must be interpreted together with its expected corrections, not automatically as failure of chiral symmetry.

On the lattice, a variational basis containing both light-cloud and internal-diquark excitations is needed to test the source assignment. Where kaon channels are open, finite-volume levels must be analyzed with the corresponding hadron-hadron channels before an excitation can be identified with a resonance. Studying several heavy masses helps separate a light-cloud excitation from $1/m_Q$ hyperfine effects and finite diquark size. Varying light masses probes the response of the parity gap to chiral dynamics. Since the physical strange quark is not massless, a claim of a chiral-limit relation requires an explicit light-flavor extrapolation or effective-theory analysis.

A complementary lattice calculation would determine the axial-current matrix element between the proposed opposite-parity states. This measures the strength of the axial transition, beyond the information contained in their masses. In the chiral-partner framework, a Goldberger--Treiman relation connects the parity gap, the axial transition strength and the coupling to the light pseudoscalar meson~\cite{NRZ2004,BEH}, with corrections away from the chiral and heavy-source limits. Calculating these quantities independently, with common field and flavor normalizations, tests that connection as the quark masses vary. For the strange isospin-singlet baryons, the direct isovector matrix element vanishes in the isospin limit: the test should use the appropriate flavor-octet current or related nonstrange channels, with the isospin-breaking factor for physical $\pi^0$ emission treated separately. At finite temperature, thermal broadening also requires care in interpreting parity-projected correlators. These comparisons test the chiral-partner description without excluding possible molecular or coupled-channel contributions.

\section{Conclusions}\label{sec:conclusions}
The three LHCb observations of 2026 make a common-light-cloud comparison experimentally concrete. The successful Ma--Harada ground-state mass prediction supports a compact, internally S-wave charm source, while leaving its radius and excited-state structure to further tests. Combining the measured masses with leading heavy-quark--diquark symmetry and meson-like parity gaps yields a consistent strange baryon spectroscopy program.

The principal parameter-insensitive results are the parity-resolved $3/4$ hyperfine relations, the relation between mesonic and baryonic differences of parity-sector hyperfine splittings, equal reduced single-pion widths, the E1 ratio $1:8:4:5$, and the spectator-limit kaon selection rule. In the mass relation, the exact identity $\Delta_{3/2}-\Delta_{1/2}=\delta_--\delta_+$ is distinguished from the approximate HDS prediction $\mathcal D_Q^{\rm bar}\simeq(3/4)\mathcal D_Q^{\rm mes}$. Each result has a stated symmetry limit; none supplies a model-independent absolute total width. The reference charm partners lie near $4.075$ and $4.183\GeV$, while bottom masses are specified relative to their ground state. The $10.27\GeV$ bottom ground mass used for illustration is motivated by consistent charm-calibrated Karliner--Rosner and lattice estimates; the primary mass predictions of this paper remain the splittings. The upper charm partner is above $\Xi_{cc}\bar K$ threshold and is protected at leading order by its light-cloud quantum numbers. Complete bottom kaon closure remains conditional on the strange--nonstrange flavor splitting.

The predicted strange doubly heavy baryons form a particularly narrow spectrum. Our benchmarks give combined pionic and electromagnetic widths of approximately $0.05$--$0.24\MeV$ for the negative-parity partners, substantially below the widths of many strongly decaying singly heavy excitations~\cite{PDG}. The positive-parity $J=\threehalf$ spin partners are narrower still, with estimated radiative widths near $8.4\keV$ for charm and $144\,\mathrm{eV}$ for bottom. This suppression reflects kaon-threshold restrictions, isospin violation in single-pion emission, and heavy-spin selection rules. The absolute values remain sensitive to electromagnetic matrix elements; additional kaon decays induced by mixing or heavy-spin breaking can broaden the upper charm partner.

The BEH- and HRS-matched pion estimates differ appreciably, but the common electromagnetic benchmark often dominates the included widths. The totals therefore do not by themselves distinguish the two chiral realizations. The mixed-flavor $bcs$ case additionally requires the heavy-heavy spin coupling and coherent mixing. Measurements of the spin partners, photon partial widths and axial matrix elements would convert the present conditional estimates into increasingly constrained tests.

\begin{acknowledgments}
M.A.N. gratefully acknowledges the inspiring discussions during the CERN workshop \href{https://indico.cern.ch/event/1718834/}{\emph{Implications and prospects of exotic hadrons at LHCb}} (October 4--7, 2026), which contributed substantially to shaping this work. In particular, we thank Liupan An, Marco Pappagallo, Ivan Polyakov, and Misha Mikhasenko for discussions of the experimental status; Marek Karliner for discussions of chiral doubling and quark models; and Luka Leskovec and Daniel Mohler for discussions of lattice calculations. We also thank Alessandro Pilloni and Manuel Pavón Valderrama for helpful discussions.

This work is supported in part by the U.S. Department of Energy under Contract No.~DE-FG02-88ER40388. M.A.N. is supported by the Priority Research Area DigiWorld under the Strategic Programme Excellence Initiative at the Jagiellonian University and the Polish National Centre of Science Grant UMO-2017/27/B/ST2/01139.

AI tools (ChatGPT Pro 5.6 Astra) were used for numerical cross-checks and to complete and update the bibliography. The final version was also checked with these tools for grammar and typographical errors. The conceptual analysis is solely due to the authors and follows the presentation at the aforementioned CERN workshop.
\end{acknowledgments}
\paragraph*{Data availability.} The numerical inputs, channel-resolved results and reproducible Python scripts accompany this article in the source package~\cite{SM}. All derivations used in the quoted benchmarks are included in the article and its appendices.

\appendix
\section{Experimental mass conventions}\label{app:experiment}

The $\Omega_{cc}^{+}$ mass uncertainties are $1.0$ (statistical), $0.2$ (systematic), $0.4$ (lifetime) and $0.6\MeV$ (external inputs)~\cite{LHCb}. The $\Xi_{cc}^{+}$ result is $3619.97\pm0.83\pm0.26{}^{+1.90}_{-1.30}\MeV$~\cite{Xi2026}. The arithmetic difference with the precision $\Xi_{cc}^{++}$ reference is $104.35\MeV$, with about $1.27\MeV$ uncertainty if correlations are neglected. Its difference from Ma--Harada's $104.9\MeV$ is $-0.55\MeV$. Their quoted $0.3\MeV$ uncertainty comes from meson inputs, not the truncation of the baryon effective theory.

Using both charge states, the unweighted isospin-average central mass is $3620.76\MeV$ and $M_\Omega-M_{\Xi,\mathrm{av}}=105.14\MeV$. Ma--Harada used the neutral-$D$ light-flavor convention; comparing that number to an isospin average is indicative, not a precision fit. Electromagnetic effects, common external inputs and correlations must be controlled before assigning significance to MeV-scale residuals. The published correlated LHCb isospin splitting, $-1.77\MeV$~\cite{Xi2026}, should be used instead of subtracting central values from different input sets. This is why the new $\Xi_{cc}^{+}$ is important for completing and checking the flavor multiplet, while the historical $\Omega_{cc}$ prediction is most directly tested with its original $\Xi_{cc}^{++}$ reference.

At leading spin-independent order, HDS gives $\bar M_{\Omega}-\bar M_{\Xi}\simeq\bar M_{H_s}-\bar M_{H_n}$. Incorporating the $3/4$ hyperfine relation when converting to spin-half masses gives
\begin{align}\label{eq:lightflavor}
 M_{\Omega,1/2}-M_{\Xi,1/2}
 &\simeq\bar M_{H_s}-\bar M_{H_n}-\half(\delta_{H_s}-\delta_{H_n})\nonumber\\
 &=\frac{3(P_s-P_n)+(V_s-V_n)}4 .
\end{align}
Here $P,V$ are pseudoscalar and vector meson masses, and the relation refers to matched light flavors with electromagnetic corrections understood. A declared legacy neutral-meson input set gives $103.97\MeV$ on the right-hand side. Its proximity to $104.35\MeV$ is a useful additional check, but cancellation of uncalculated corrections cannot be excluded. It is not a determination of $r_{cc}$.

\section{Hyperfine algebra and unequal parity splittings}\label{app:masses}
Write the heavy-light meson spin interaction as $A_Q\bm s_Q\cdot\bm j_\ell$ with $j_\ell=\half$. Its $J=1$ and $J=0$ expectation values differ by one, giving $\delta_M=A_Q$. For a compact color-antitriplet pair of identical quarks, the matched baryon interaction is $(A_Q/2)\bm S_d\cdot\bm j_\ell$. Since
\begin{equation}
\begin{aligned}
 \langle\bm S_d\cdot\bm j_\ell\rangle_J
 &=\frac{J(J+1)-S_d(S_d+1)-j_\ell(j_\ell+1)}2\\
 &=\begin{cases}-1,&J=\half,\\+\half,&J=\threehalf,\end{cases}
\end{aligned}
\end{equation}
one obtains $(A_Q/2)(3/2)=3\delta_M/4$. The color normalization in the magnetic matching is essential; angular momentum alone does not fix it~\cite{BVR,FM}.

For arbitrary $\delta_\pm$, the centroid parametrization gives
\begin{align}
 M_{1/2}^\pm&=\overline M_\pm-\frac23\delta_\pm,&
 M_{3/2}^\pm&=\overline M_\pm+\frac13\delta_\pm,\\
 \Delta_{1/2}&=\Delta_\chi-\frac23(\delta_- -\delta_+),&
 \Delta_{3/2}&=\Delta_\chi+\frac13(\delta_- -\delta_+).\label{eq:rect}
\end{align}
Subtracting the two gaps in Eq.~\eqref{eq:rect} gives $\Delta_{3/2}-\Delta_{1/2}=\delta_--\delta_+=\mathcal D_Q^{\rm bar}$, consistently with Eq.~\eqref{eq:defect}. One cannot impose independent unequal parity hyperfine intervals and identical spin-resolved chiral gaps simultaneously. The rounded reference in Table~\ref{tab:masses} sets $\delta_-=\delta_+$ and hence $\mathcal D_Q^{\rm bar}=0$.

As a refinement, the rounded meson intervals $\delta_H=143.8\MeV$ and $\delta_G=141.8\MeV$ give $\mathcal D_c^{\rm mes}=-2.0\MeV$. Leading HDS then gives $\delta_+^{cc}=107.85\MeV$, $\delta_-^{cc}=106.35\MeV$ and $\mathcal D_c^{\rm bar}=-1.50\MeV$. Holding the specified centroid gap at $349\MeV$, Eq.~\eqref{eq:rect} yields
\[
\begin{aligned}
 \Delta_{1/2}&=349-\frac23(-1.50)=350.0\MeV,\\
 \Delta_{3/2}&=349+\frac13(-1.50)=348.5\MeV.
\end{aligned}
\] The negative-parity masses become $4075.90$ and $4182.25\MeV$; the A included widths change to $84.09$ and $239.78\keV$. This small shift is much smaller than the E1 uncertainty. It does not change the threshold conclusion. Matching the exact measured \emph{meson} centroid would be a different numerical input from the stipulated $349\MeV$.

For completeness, transfer the centroid of the declared legacy meson set
\begin{equation}
\begin{gathered}
 (M_{D_s},M_{D_s^*},M_{D_{s0}},M_{D_{s1}})\\
 =(1968.35,2112.20,2317.70,2459.50)\MeV.
\end{gathered}
\end{equation}
This set gives $\delta_H=143.85\MeV$, $\delta_G=141.80\MeV$, $d_0=349.35\MeV$ and $d_1=347.30\MeV$. Thus $\mathcal D_c^{\rm mes}=-2.05\MeV$, $\mathcal D_c^{\rm bar}\simeq-1.5375\MeV$ and the transferred centroid gap is $\Delta_\chi=347.8125\MeV$. Equation~\eqref{eq:resolved} then gives $\Delta_{1/2}=348.8375\MeV$ and $\Delta_{3/2}=347.3000\MeV$. Applying $3/4$ to each parity hyperfine interval yields
\begin{equation}
\begin{gathered}
 (M_{B_{1/2}},M_{B_{3/2}},M_{X_{1/2}},M_{X_{3/2}})\\
 =(3725.90,3833.7875,4074.7375,4181.0875)\MeV.
\end{gathered}
\end{equation}
This is a distinct input prescription, reproduced by \texttt{experimental\_checks.py}, not an additional correction to the rounded reference table. The separate 2026 LHCb input $\mathcal D_c^{\rm mes}=-2.7\pm0.7\MeV$ instead gives $\mathcal D_c^{\rm bar}\simeq-2.03\pm0.53\MeV$, with input error only, as explained below Eq.~\eqref{eq:defect}~\cite{Bs2026}. These values differ because their mesonic inputs differ. The updated difference alone does not specify the centroid gap. We do not combine it with the legacy mass set into a precision fit or apply it as a correction to Table~\ref{tab:masses}.

\subsection{Absolute-mass illustration and binding increments}
The centroid inferred from the measured charm spin-half mass and the assumed hyperfine relation is $\overline M_{\Omega_{cc}}=3725.90+(2/3)107.85=3797.80\MeV$. The bottom spin-half mass lies $24.25\MeV$ below its centroid. These two conversions yield Eq.~\eqref{eq:absolute-bottom}; replacing a charm spin-half mass by a bottom spin-half mass without them would omit a $47.65\MeV$ term.

In the notation of the Karliner--Rosner calculation~\cite{KarlinerRosner2018}, let $m_{QQ}$ include the internal heavy-heavy binding and hyperfine energy, and let $B((QQ)s)>0$ denote the additional attractive binding to the strange quark. The strange constituent mass and any common baryon junction term cancel within each prescription. Thus
\begin{equation}
 \Delta_{\rm cent}=(m_{bb}-m_{cc})-[B((bb)s)-B((cc)s)].
\end{equation}
Their separate meson/baryon constituent-mass prescription gives $(m_{cc},m_{bb})=(3306.2,9813.4)\MeV$ and $(B((cc)s),B((bb)s))=(109.4,129.4)\MeV$. The universal-mass prescription gives $(3204.1,9718.9)\MeV$ and $(81.6,98.4)\MeV$, respectively. Consequently,
\begin{align}
 \Delta_{\rm cent}^{\rm separate}&=6507.2-20.0=6487.2\MeV,\\
 \Delta_{\rm cent}^{\rm universal}&=6514.8-16.8=6498.0\MeV.
\end{align}
Substitution into Eq.~\eqref{eq:absolute-bottom} gives $10260.75$ and $10271.55\MeV$. This procedure transfers the model's spin-independent heavy-flavor increment and recalibrates its common mass offset to the charm centroid. It does not constitute a refit of the quark model or a replacement for a calculation of finite-size and heavy-mass corrections. The published binding increments are retained as external model inputs; the constituent mass adopted for our electromagnetic operators does not redefine them. We retain the lattice value as a separate comparison rather than form a statistical combination without controlled theory correlations. The numerical script records these inputs and calculations alongside the illustrative bottom mass.

\section{From meson amplitudes to baryon widths}\label{app:transfer}
BEH~\cite{BEH} already outlines the doubly heavy baryon extension in its Sec.~VI. Our adaptation of BEH and HRS~\cite{HRS} transfers reduced light-cloud amplitudes, with the spectator and external-state factors treated separately. The underlying source correspondence is between $\bar Qs$ and $(QQ)_{\bar{\mathbf3}}s$, using charge-conjugate mesons where appropriate. Their heavy sources have the same color representation but baryon numbers $-1/3$ and $2/3$, respectively. This correspondence relates matrix elements in distinct sectors; it is not a meson-to-baryon conversion.

Both $B_J$ and $X_J$ carry baryon number one. In a canonically normalized heavy-field basis the leading spin-diagonal pion vertex can be written schematically as
\begin{equation}\label{eq:baryon-number-vertex}
 \mathcal L_\pi=g_\pi\sum_{J,m}B_{Jm}^{\dagger}X_{Jm}
 v\!\cdot\!\partial\pi^0+\mathrm{h.c.},
\end{equation}
where $g_\pi$ includes the flavor factor, inverse pion decay constant and isospin-breaking mixing. Under baryon-number rotations, $B_{Jm},X_{Jm}\to e^{i\alpha}(B_{Jm},X_{Jm})$; the bilinear is invariant. Thus every retained strong or electromagnetic transition has baryon number $1\to1+0$. There is no extra multiplicity from baryon number or from the two heavy quarks: the pion probes one light cloud.

For a compact static source, HDS relates the leading meson and baryon light-cloud couplings~\cite{HuMehen}. At this order the scalar pion operator has unit spin-diagonal strength, whereas vector transitions require the recoupling factors in Eq.~\eqref{eq:robustweights}. The practical pion-width matching is
\begin{equation}\label{eq:transfer-width}
 \Gamma_B=\Gamma_M^{\rm ref}\,
 \frac{\Phi(M_i^{B},M_f^{B})}{\Phi(M_i^{M},M_f^{M})},
\end{equation}
with $\Phi$ defined in Eq.~\eqref{eq:pi}. The leading amplitude contains $2\sqrt{M_iM_f}$ from external-state normalization, leaving $M_f/M_i$ after two-body phase space. At fixed gap, $M_f/M_i\to1$ as the heavy mass increases; there is no universal factor-of-four suppression. Exact baryon masses enter the retained pion, photon and dipion kinematics. The hyperfine $3/4$ relation sets the spectrum, not an overall decay-width factor.

The larger source mass also enters the electromagnetic operators: the constituent benchmark uses $M_d=2m_Q$ and $e_d=2e_Q$ in Eq.~\eqref{eq:charge}, and sums the two heavy-quark magnetic moments. These changes are separate from spin recoupling. The HRS meson-field normalization parameter $M_Q=1974\MeV$ must not simply be doubled; Appendix~\ref{app:normalization} explains its conversion to our matched coefficient.

Dynamical mass effects remain approximate. Replacing the antiquark by a diquark changes the reduced mass from $m_sm_Q/(m_s+m_Q)$ to $2m_sm_Q/(m_s+2m_Q)$, and can therefore change radial overlaps and transition form factors. We do not recompute baryon wavefunctions, internal-diquark binding corrections to the electromagnetic constituent mass, or a complete set of finite-mass and finite-size operators. The common $R=2.5\GeV^{-1}$ overlap is a declared benchmark with a sensitivity study. At $m_s=0.323\GeV$, the strange-quark--diquark reduced masses are approximately $292.2$, $312.3$ and $307.2\MeV$ for $ccs$, $bbs$ and $bcs$, respectively. Changing the constituent mass can change the radial wavefunctions, but the overlap is not refitted here. The quoted results are consequently BEH/HRS-matched baryon estimates; the HRS-matched hadronic contribution is combined with the same separately specified electromagnetic model used in A.

\section{Single-pion normalization and literature audit}\label{app:normalization}
For canonically normalized heavy states, choose the spin-diagonal reduced amplitude
\begin{equation}
 \mathcal M=2\sqrt{M_iM_f}\,g_\pi E_\pi\,\chi_f^\dagger\chi_i.
\end{equation}
A spin average and two-body phase space give
\begin{equation}
 \Gamma=\frac{p}{8\pi M_i^2}\overline{|\mathcal M|^2}
 =\frac{|g_\pi|^2}{2\pi}\frac{M_f}{M_i}E_\pi^2p,
 \qquad C_\pi=\frac{|g_\pi|^2}{2\pi}.
\end{equation}
An $\eta$--$\pi^0$ mixing factor can be included in $g_\pi$. Different definitions of that factor, flavor generators and heavy-field normalization must not be combined as though they were the same coupling.

In particular, HRS~\cite{HRS} Eq.~(6.9) is
\begin{equation}\label{eq:hrspaper}
\Gamma=\frac{p}{2\pi}\left[\frac{k_{\rm HRS}}{F_\pi}\frac{M_Q}{M_i}E_\pi\Delta_{\pi^0\eta}\right]^2.
\end{equation}
Here $k_{\rm HRS}=0.59$, $F_\pi=92.42\MeV$ and $\Delta_{\pi^0\eta}=-5.32\times10^{-3}$. We attach the subscript HRS to distinguish this dimensionless coupling from the photon momentum. Its $M_Q=1974\MeV$ is the chosen nonstrange heavy-meson spin average, not a constituent charm mass or a baryon mass. The rounded table inputs give about $4.18$ and $3.76\keV$, consistent with the quoted $4.17$ and $3.75\keV$. Replacing this field-normalization parameter by a guessed baryon mass would introduce an arbitrary extrapolation. Matching instead gives
\begin{equation}
 C_\pi^{\rm B}\simeq\frac{k_{\rm HRS}^2\Delta_{\pi^0\eta}^2}{2\pi F_\pi^2}
 \frac{M_Q^2}{M_{D_{s0}}M_{D_s}}.
\end{equation}
The last factor is about $0.854$. Using the axial HRS channel as a separate calibration would reduce $C_\pi^{\rm B}$ by roughly $12\%$. This spread is a finite-mass matching ambiguity, not a new spin-symmetry prediction. At the level of accuracy of this study, one reference channel is preferable to inserting inconsistent spin-dependent normalizations.

BEH~\cite{BEH} explicitly uses $f_\pi=93.3\MeV$ after its Eq.~(9). Its nonstrange neutral-pion vertex, Eq.~(32), together with the canonical heavy-field normalization in its Appendix~A, gives $g_\pi=G_A/(2f_\pi)$ in our notation, before isospin mixing. Therefore
\begin{equation}\label{eq:beh-normalization}
 \Gamma_{\pi^0}=\frac{G_A^2}{8\pi f_\pi^2}\frac{M_f}{M_i}E_\pi^2p.
\end{equation}
The static prescription $M_f/M_i\to1$, $E_\pi\to\Delta M$, with $\Delta M=349\MeV$ and $p=297\MeV$, gives $165G_A^2\MeV$, close to BEH's quoted $164G_A^2\MeV$. Its printed Eq.~(33) instead has $1/(4\pi f_\pi^2)$ and gives $331G_A^2\MeV$ with these inputs. This indicates an apparent factor-of-two prefactor misprint; it does not establish that BEH changed decay-constant conventions. Indeed, $1/(8\pi f_\pi^2)=1/(4\pi f_{132}^2)$ for $f_{132}=\sqrt2f_\pi$, so a consistent convention change leaves the width invariant. Multiplying the quoted $164\MeV$ by $[(1/2)/43.7]^2$ reproduces $21.5\keV$. We retain this published numerical benchmark as the A matching condition. The apparent prefactor error consequently changes none of our tabulated predictions or conclusions.

Separately, choosing the coefficient $G_A^2\delta_{\eta\pi}^2/(8\pi f_\pi^2)$ with $\delta_{\eta\pi}=(1/2)/43.7$ gives $5.984\times10^{-4}\GeV^{-2}$, about $25\%$ below Eq.~\eqref{eq:match}. This alternative inserts exact recoil kinematics without recalibrating to the quoted static meson width. It gives baryon pion widths of approximately $19.1$ and $18.6\keV$ for charm and bottom, rather than $25.3$ and $24.7\keV$; the included A widths decrease by only about $6\keV$. Conversely, removing HRS's meson field-normalization factor before transfer gives $1.836\times10^{-4}\GeV^{-2}$, about $18\%$ above our matched B value. These are alternative finite-mass matching prescriptions, not physical effects of changing the convention for $f_\pi$. They leave the mass relations, reduced-width ratios, threshold arguments and predicted pattern of narrow states intact.

\section{Spin recoupling and electromagnetic operators}\label{app:angular}
\subsection{Wigner--Eckart convention and spectator-spin reduction}
Let $T_q^{(k)}$ be a spherical tensor of rank $k$ acting only on the light cloud, with the heavy-diquark spin $S$ unchanged. We couple the light spin first and use Condon--Shortley phases:
\begin{equation}\label{eq:coupledstates}
 |(jS)JM\rangle=\sum_{m,m_S}
 C^{JM}_{jm,Sm_S}|jm\rangle|Sm_S\rangle.
\end{equation}
The parity labels on the light states are implicit; they affect the reduced matrix element but not this spin algebra. Define
\begin{equation}\label{eq:lightWE}
\begin{aligned}
 \langle j_fm_f|T_q^{(k)}|j_im_i\rangle
 &=(-1)^{j_f-m_f}
 \begin{pmatrix}j_f&k&j_i\\-m_f&q&m_i\end{pmatrix}
 t_{fi}^{(k)},\\
 t_{fi}^{(k)}&\equiv\langle j_f\Vert T^{(k)}\Vert j_i\rangle.
\end{aligned}
\end{equation}
There is no extra factor $1/\sqrt{2j_f+1}$ multiplying the $3j$ symbol in this convention. Expanding both coupled states and using the spectator overlap $\langle Sm_S'|Sm_S\rangle=\delta_{m_S'm_S}$ gives
\begin{equation}\label{eq:CGsum}
\begin{aligned}
 &\langle(j_fS)J_fM_f|T_q^{(k)}|(j_iS)J_iM_i\rangle\\
 &\quad=t_{fi}^{(k)}\sum_{m_f,m_i,m_S}
 C^{J_fM_f}_{j_fm_f,Sm_S}C^{J_iM_i}_{j_im_i,Sm_S}\\
 &\qquad\qquad\times(-1)^{j_f-m_f}
 \begin{pmatrix}j_f&k&j_i\\-m_f&q&m_i\end{pmatrix}.
\end{aligned}
\end{equation}
The recoupling identity for this sum is the total-spin Wigner--Eckart theorem,
\begin{equation}\label{eq:totalWE}
\begin{aligned}
 &\langle(j_fS)J_fM_f|T_q^{(k)}|(j_iS)J_iM_i\rangle\\
 &\quad=(-1)^{J_f-M_f}
 \begin{pmatrix}J_f&k&J_i\\-M_f&q&M_i\end{pmatrix}
 \mathcal T_{fi}^{(k)},\\
 \mathcal T_{fi}^{(k)}
 &=(-1)^{j_f+S+J_i+k}\sqrt{(2J_f+1)(2J_i+1)}\\
 &\quad\times
 \begin{Bmatrix}j_f&J_f&S\\J_i&j_i&k\end{Bmatrix}
 t_{fi}^{(k)}.
\end{aligned}
\end{equation}
This fixes the phase as well as the magnitude in the stated coupling convention. The phase drops out of each separate partial width, but must be retained when coherently combining amplitudes for mixed states.

The relevant strength is averaged over initial polarizations and summed over final polarizations and tensor components. Orthogonality of the $3j$ symbols yields
\begin{equation}\label{eq:averagedstrength}
\begin{aligned}
 \mathcal B_{J_iJ_f}^{(k)}
 &\equiv\frac{1}{2J_i+1}\sum_{M_i,M_f,q}
 |\langle J_fM_f|T_q^{(k)}|J_iM_i\rangle|^2\\
 &=\frac{|\mathcal T_{fi}^{(k)}|^2}{2J_i+1}\\
 &=(2J_f+1)
 \begin{Bmatrix}j_f&J_f&S\\J_i&j_i&k\end{Bmatrix}^{\!2}
 |t_{fi}^{(k)}|^2.
\end{aligned}
\end{equation}
The analogous isolated light-cloud strength is
$\mathcal B_\ell^{(k)}=|t_{fi}^{(k)}|^2/(2j_i+1)$.
Their ratio is therefore
\begin{equation}\label{eq:6j}
\begin{aligned}
 w_{J_iJ_f}^{(k)}&\equiv
 \frac{\mathcal B_{J_iJ_f}^{(k)}}{\mathcal B_\ell^{(k)}}\\
 &=(2j_i+1)(2J_f+1)
 \begin{Bmatrix}j_f&J_f&S\\J_i&j_i&k\end{Bmatrix}^{\!2}.
\end{aligned}
\end{equation}
In particular, the factor $2J_i+1$ from the squared reduced matrix element cancels against the initial-spin average. The remaining final-spin multiplicity $2J_f+1$ is essential.

\subsection{Explicit evaluation of the four spin-channel weights}
For a rank-one transition between the two $j_\ell=\half$ light clouds around a spin-one diquark,
\begin{equation}\label{eq:weightspecial}
 w_{J_iJ_f}=2(2J_f+1)
 \begin{Bmatrix}\half&J_f&1\\J_i&\half&1\end{Bmatrix}^{\!2},
 \qquad J_i,J_f=\half,\threehalf.
\end{equation}
For completeness, the $6j$ symbols can be evaluated directly with the Racah factorial sum~\cite{DLMF34}. Define the triangle coefficient
\begin{equation}\label{eq:trianglefactor}
 \Delta(a,b,c)=\sqrt{\frac{(a+b-c)!(a-b+c)!(-a+b+c)!}
 {(a+b+c+1)!}}
\end{equation}
and the products and summation bounds
\begin{equation}\label{eq:Racahbounds}
\begin{aligned}
 P_\Delta&=\Delta(a,b,c)\Delta(a,e,f)
              \Delta(d,b,f)\Delta(d,e,c),\\
 (x_1,x_2,x_3,x_4)&=(a+b+c,a+e+f,\\
 &\hspace{3.1em}d+b+f,d+e+c),\\
 (y_1,y_2,y_3)&=(a+b+d+e,a+c+d+f,\\
 &\hspace{3.1em}b+c+e+f).
\end{aligned}
\end{equation}
Then
\begin{equation}\label{eq:Racahsum}
 \begin{Bmatrix}a&b&c\\d&e&f\end{Bmatrix}
 =P_\Delta\sum_{z=\max x_p}^{\min y_q}
 \frac{(-1)^z(z+1)!}
 {\prod_{p=1}^{4}(z-x_p)!\prod_{q=1}^{3}(y_q-z)!}.
\end{equation}
Only integer $z$ with nonnegative factorial arguments contribute. With $(a,b,c;d,e,f)=(\half,J_f,1;J_i,\half,1)$, each of the four sums contains exactly one term. The complete evaluations are
\begin{center}
\setlength{\tabcolsep}{4pt}
\renewcommand{\arraystretch}{1.5}
\begin{tabular}{ccccc}
\toprule
$(J_i,J_f)$ & $P_\Delta$ & $z$ & Summand & $6j$\\
\midrule
$(\half,\half)$ & $1/36$ & $2$ & $6$ & $1/6$\\
$(\half,\threehalf)$ & $1/72$ & $3$ & $-24$ & $-1/3$\\
$(\threehalf,\half)$ & $1/72$ & $3$ & $-24$ & $-1/3$\\
$(\threehalf,\threehalf)$ & $1/(72\sqrt{10})$ & $4$ & $60$ & $\sqrt{10}/12$\\
\bottomrule
\end{tabular}
\end{center}
For example, in the first row all four triangle coefficients are $1/\sqrt6$ and the factorial summand is $3!=6$, giving $6/36=1/6$. In the last row the triangle coefficients are $1/\sqrt{12}$, $1/\sqrt6$, $1/\sqrt{60}$ and $1/\sqrt{12}$; the only nontrivial denominator factorial is $2!$, so the summand is $5!/2!=60$.

Substitution into Eq.~\eqref{eq:weightspecial} gives every entry separately:
\begin{equation}\label{eq:fourweights}
\begin{aligned}
 w_{\half\,\half}&=2\times2\left(\frac16\right)^2=\frac19,\\
 w_{\half\,\threehalf}&=2\times4\left(-\frac13\right)^2=\frac89,\\
 w_{\threehalf\,\half}&=2\times2\left(-\frac13\right)^2=\frac49,\\
 w_{\threehalf\,\threehalf}&=2\times4\left(\frac{\sqrt{10}}{12}\right)^2
 =\frac59.
\end{aligned}
\end{equation}
Thus Eq.~\eqref{eq:robustweights} is
\begin{equation}\label{eq:weightmatrixderived}
 (w_{J_iJ_f})=\frac19\begin{pmatrix}1&8\\4&5\end{pmatrix},
\end{equation}
with initial spins labeling rows and final spins labeling columns. The unequal off-diagonal entries arise from the final-spin multiplicity after the initial-spin average; their $6j$ symbols are equal. Equivalently,
$(2J_i+1)w_{J_iJ_f}=(2J_f+1)w_{J_fJ_i}$ for these weights.

\subsection{Normalization checks and the electric-dipole prefactor}
Orthogonality of the $6j$ symbols gives
\begin{equation}\label{eq:weightsum}
\begin{aligned}
 \sum_{J_f}(2J_f+1)
 \begin{Bmatrix}j_f&J_f&S\\J_i&j_i&k\end{Bmatrix}^{\!2}
 &=\frac{1}{2j_i+1},\\
 \sum_{J_f}w_{J_iJ_f}^{(k)}&=1.
\end{aligned}
\end{equation}
The two explicit row sums are $(1+8)/9=(4+5)/9=1$. They exhaust the light-cloud strength for either initial spin, before channel-dependent photon energies are included. The weights are consequently not physical branching fractions when those energies differ. Dividing the E1 widths by their individual $k_{if}^3$ factors gives the ratio $1:8:4:5$ of Eq.~\eqref{eq:e1ratio}, under the common-overlap assumption.

For a scalar light operator, $k=0$, the $3j$ triangle condition requires $J_i=J_f$; the single allowed weight is unity. This proves the leading single-pion spin relation. For a heavy meson, setting $S=\half$ in Eq.~\eqref{eq:6j} instead gives $w_{0\to1}=1$, $w_{1\to0}=1/3$, $w_{1\to1}=2/3$ and $w_{0\to0}=0$. Hence a baryon rank-one width calibrated to a meson $1\to0$ width carries the factor $3w$ used below.

To derive the electric prefactor, remove the center of mass from $e(e_s\bm r_s+e_d\bm r_d)$. The residual coordinate has coefficient Eq.~\eqref{eq:charge}. The spin-averaged orbital strength for $L_i=1\to L_f=0$ is $R^2/3$. Multiplying it by the dipole emission factor $4\alpha k^3/3$ and the fraction $w$ gives Eq.~\eqref{eq:E1}. Here $k$ is the photon momentum, as in the main text. The meson specialization is equivalent to the BEH coefficient $(4\alpha/27)e_{\rm eff}^2k^3(2J_f+1)R^2S_{if}$. This fixes the normalization independently of the baryon $6j$ evaluation.

\subsection{Magnetic-dipole operators}

For magnetic transitions define $\mu_i=e_i/(2m_i)$. In a spin-one diquark,
\begin{equation}
 \bm\mu_d=(\mu_Q+\mu_{Q'})\bm S_d\equiv a_d\bm S_d,
 \qquad a_S=2\mu_s.
\end{equation}
For the relative P wave, the orbital magnetic coefficient is
\begin{equation}
 \mu_L=\mu_s\frac{M_d}{M_d+m_s}
 +\frac{e_d}{2M_d}\frac{m_s}{M_d+m_s}.
\end{equation}
Projection onto $L=1,s=\half,j_\ell=\half$ gives $\bm L\to(4/3)\bm j_\ell$ and $\bm s\to-(1/3)\bm j_\ell$, and therefore
\begin{equation}
 a_P=\frac43\mu_L-\frac23\mu_s.
\end{equation}
Between distinct total-spin states, the matrix element of $\bm J=\bm S_d+\bm j_\ell$ vanishes; the transition consequently depends on $(a_{S,P}-a_d)\bm j_\ell$. The averaged light-spin vector strength for $J_i=\threehalf\to J_f=\half$ is $1/3$, producing Eq.~\eqref{eq:M1}. Setting an S-wave magnetic coefficient in the negative-parity pair would omit its orbital contribution.

\section{One-dimensional dipion integrals}\label{app:dipion}
\subsection{Scalar channel}
Let $s=(p_1+p_2)^2$, $q(s)=p(M_i,M_f,\sqrt s)$ and $\beta(s)=\sqrt{1-4m_\pi^2/s}$. The scalar amplitude has a rank-one recoil factor and the shape
\begin{equation}
 F_S(s;m_\pi)=(s-4m_\pi^2)
 \left(\frac{1}{s-m_{\sigma0}^2}+\frac{1}{s-m_{\sigma8}^2}\right).
\end{equation}
Factorizing three-body phase space into baryon--dipion and two-pion phase spaces supplies $q\beta\,\dd s$; squaring the recoil factor supplies $q^2$. In our matched normalization, define
\begin{equation}\label{eq:Is}
\begin{aligned}
 I_S(M_i,M_f;m_\pi)&=\frac{M_f}{M_i}\\
 &\quad\times\int_{4m_\pi^2}^{(M_i-M_f)^2}\dd s\,q^3\beta|F_S(s;m_\pi)|^2,\\
 \mathcal I_S&=I_S(m_{\pi^0})+2I_S(m_{\pi^+}).
\end{aligned}
\end{equation}
It vanishes when the integration interval is empty. With reference masses $(M_i^D,M_f^D)=(2.4595,1.96835)\GeV$, the baryon contribution is
\begin{equation}\label{eq:gs}
 \Gamma_{\pi\pi}^S(J_i,J_f)=3w_{J_iJ_f}(4.2\keV)
 \frac{\mathcal I_S(M_i,M_f)}{\mathcal I_S(M_i^D,M_f^D)}.
\end{equation}
The individual neutral and charged widths follow by replacing the numerator by $I_S(m_{\pi^0})$ and $2I_S(m_{\pi^+})$. This enforces the $2:1$ charged/neutral ratio in the isospin and equal-mass limit. For the reference masses the charged-to-neutral integral ratio before multiplication by two is about $0.856$.

BEH~\cite{BEH} lists separate neutral and charged numbers $2.3$ and $1.9\keV$. Using them as two unrelated normalizations would not reproduce the isoscalar multiplicity of the explicitly adopted amplitude in the equal-mass limit. We retain only their summed $4.2\keV$ scale and state the adaptation in Eq.~\eqref{eq:gs}. This is one reason to regard the dipion estimate as illustrative. Close to threshold, the explicit $(s-4m_\pi^2)^2$ factor makes it much more suppressed than a constant-amplitude phase-space estimate.

\subsection{HRS vector-mixing channel}
For the vector term take
\begin{align}
 F_V(s)&=\frac{m_\rho^2\Pi_{\phi\rho}}
 {(s-m_\phi^2)(s-m_\rho^2)},\\
 m_\rho&=0.7711\GeV,\qquad m_\phi=1.019456\GeV,\nonumber\\
 \Pi_{\phi\rho}&=0.000530\GeV^2.
\end{align}
The dipion masses are far below the vector poles; widths in the propagators are omitted consistently with the reference calculation. HRS~\cite{HRS} Eq.~(6.26) contains the Dalitz kernel
\begin{equation}
 K(s,t)=s-4m_\pi^2+
 \frac{[s-M_i^2-M_f^2-2m_\pi^2+2t]^2}{4M_i^2},
\end{equation}
where $t=(p_f+p_1)^2$. At fixed $s$,
\begin{equation}
 t_0=\frac{M_i^2+M_f^2+2m_\pi^2-s}{2},\qquad
 t_\pm=t_0\pm M_iq\beta .
\end{equation}
Writing $x=t-t_0$ reduces the kernel to $s-4m_\pi^2+x^2/M_i^2$, so the $t$ integral is elementary:
\begin{equation}\label{eq:tint}
 \int_{t_-}^{t_+}\dd t\,K=2M_iq\beta
 \left(s-4m_\pi^2+\frac{q^2\beta^2}{3}\right).
\end{equation}
After matching to the common heavy-state normalization, the required shape is
\begin{equation}\label{eq:Iv}
\begin{aligned}
 I_V(M_i,M_f)&=\frac{M_f}{M_i}\int_{4m_{\pi^+}^2}^{(M_i-M_f)^2}\dd s\,q\beta\\
 &\quad\times\left(s-4m_{\pi^+}^2+\frac{q^2\beta^2}{3}\right)|F_V(s)|^2.
\end{aligned}
\end{equation}
Then
\begin{equation}
 \Gamma_{\pi\pi}^V(J_i,J_f)=3w_{J_iJ_f}(2.13\times10^{-4}\keV)
 \frac{I_V(M_i,M_f)}{I_V(2.460,1.969)}.
\end{equation}
This is a reduced-amplitude transfer of the HRS vector result, with the same finite-mass matching limitation as the pion channel. A neutral identical-pion P wave is excluded by Bose symmetry. Both $I_S$ and $I_V$ have mass dimension five; their ratios are dimensionless. The supplied code independently checks Eq.~\eqref{eq:tint} by numerical integration over $t$.

\section{How to extend the calculation to physical \texorpdfstring{$bcs$}{bcs} states}\label{app:bc}
The two spin-half eigenvalues of Eq.~\eqref{eq:bcmat} are
\begin{equation}
 E_\pm=-\frac c4-\frac{a+b}{4}
 \pm\frac12\sqrt{\left[c-\frac{a+b}{2}\right]^2+\frac34(b-a)^2}.
\end{equation}
The mixing angle is fixed, up to basis phases, by $\tan2\theta=2H_{01}/(H_{11}-H_{00})$. Values of $c$, and potentially distinct $a,b,c$ in the two parity sectors, are required. If $U_+$ and $U_-$ diagonalize these sectors, any transition operator must be transformed as
\begin{equation}
 \mathcal A_{\rm physical}=U_+^\dagger\mathcal A_{\rm basis}U_-.
\end{equation}
For a common light-cloud S-wave pion coupling, $\mathcal A_{\rm basis}$ is proportional to the identity in heavy-spin space. Equal mixing angles then preserve diagonal pion decays; different angles generate off-diagonal channels. For E1 and scalar/vector dipion transitions, the signed reduced matrix elements in Eq.~\eqref{eq:totalWE} must be used \emph{before} squaring. The tabulated positive weights alone lose the interference information.

The $bc$ magnetic operator also contains a component absent for identical heavy quarks:
\begin{equation}
 \bm\mu_{bc}=(\mu_b+\mu_c)\bm S_{bc}
 +(\mu_b-\mu_c)(\bm s_b-\bm s_c).
\end{equation}
The second term connects $S_{bc}=0$ and $1$. It opens additional magnetic transitions between physical spin-half levels. A complete $bcs$ width prediction must include these transitions and the resulting physical thresholds. The supplied mixing calculation provides the spin energies and probabilities for illustrative $c=0,25,50,75,100\MeV$, but does not pretend that these arbitrary choices determine the spectrum.

\section{Reproducibility and limitations}\label{app:repro}
All numerical calculations use GeV internally and convert widths to keV by multiplying by $10^6$. The pion masses are $m_{\pi^0}=0.1349768$ and $m_{\pi^+}=0.13957039\GeV$; $\alpha^{-1}=137.035999084$. The supplementary script \texttt{calculate.py} requires Python, NumPy, SciPy and Matplotlib. Running it writes \texttt{results.json}, \texttt{widths.csv} and \texttt{sensitivity.pdf}. The JSON records the constituent-mass inputs, channel-resolved results, overlap variations, a parity-hyperfine refinement, the kaon mixing illustration and the mixed-flavor spin matrices. The companion \texttt{experimental\_checks.py} writes \texttt{experimental\_checks.json} with the 2026 comparisons and alternative centroid prescription. Run \texttt{build.sh} to compile the PRD article. Both figure files, \texttt{baryonic\_chiral\_rectangles\_caption.tex}, and \texttt{references.tex} must remain alongside the sources. The script \texttt{baryonic\_chiral\_rectangles.py} generates Fig.~\ref{fig:baryonic-rectangles} from \texttt{results.json}.

Internal checks verify all nonzero E1 weights and their sum rules, their meson limits, the matching conditions, closure of phase space and the independent Dalitz reduction. These checks establish the implementation of the stated model; they do not validate its assumed matrix elements against QCD. No unpublished baryon coupling has been fitted to data. Quoted mass gaps, scalar shapes, radial overlaps and the point-diquark approximation can be replaced in the code when improved inputs become available.

The calculation retains the lowest light-cloud parity doublets only. Additional internal-diquark levels, molecular components, coupled-channel dressing, higher electromagnetic multipoles and independent isospin- or heavy-spin-breaking operators may alter physical widths. In particular, radiative dominance and the upper-charm kaon contribution must be tested rather than assumed. The reliable deliverable is a transparent set of conditional estimates and scaling relations with an explicit separation between symmetry statements, adopted inputs and model-dependent matrix elements.

\end{document}